\documentclass[twocolumn]{autart}    

\usepackage{graphicx}          

\usepackage{amsmath,mathrsfs,mathtools,amssymb}
\usepackage{enumerate}
\usepackage{siunitx}

\usepackage{cite}

\usepackage{booktabs}
\usepackage{tabularx}

\begin{document}

\setlength{\abovedisplayskip}{4pt}
\setlength{\belowdisplayskip}{4pt}

\begin{frontmatter}

\title{Synergistic Potential Functions from Single Modified Trace Function on SO(3)\thanksref{footnoteinfo}} 

\thanks[footnoteinfo]{Research reported in this work was supported in part by Innovation and Technology Commission of Hong Kong (ITS/135/20, ITS/136/20, ITS/233/21, ITS/234/21), Research Grants Council (RGC) of Hong Kong (CUHK 24201219, CUHK 14217822), project BME-p7-20 of the Shun Hing Institute of Advanced Engineering, and University of Sydney-CUHK Partnership Collaboration Awards. The content is solely the responsibility of the authors and does not reflect the views of the sponsors. Corresponding author is S. S. Cheng.}

\author[MAE,MRC]{Xin Tong}
\ead{xtong@cuhk.edu.hk},
\author[MAE,MRC,CURI,SHI]{Shing Shin Cheng}
\ead{sscheng@cuhk.edu.hk}

\address[MAE]{Department of Mechanical and Automation Engineering, The Chinese University of Hong Kong, Hong Kong}
\address[MRC]{Multi-Scale Medical Robotics Center, The Chinese University of Hong Kong, Hong Kong}
\address[CURI]{CUHK T Stone Robotics Institute, The Chinese University of Hong Kong, Hong Kong}
\address[SHI]{Shun Hing Institute of Advanced Engineering, The Chinese University of Hong Kong, Hong Kong}      

\begin{abstract}
    This paper is about the construction of a family of centrally synergistic potential functions from a single modified trace function on $\mathrm{SO(3)} $. First, we demonstrate that it is possible to complete the construction through angular warping with multiple directions, particularly effective in the unresolved cases in the literature. Second, it can be shown that for each potential function in the family, there exists a subset of the family such that the synergistic gap is positive at the unwanted critical points. This allows the switching condition to be checked within the selected subsets while implementing synergistic hybrid control. Furthermore, the positive lower bound of synergistic gap is explicitly expressed by selecting a traditional warping angle function. Finally, we apply the proposed synergistic potential functions to obtain robust global attitude tracking.
\end{abstract}
  

\end{frontmatter}

\section{Introduction}

Attitude tracking control crops up everywhere in robot manipulators, aerospace as well as precision machine, and has therefore generated considerable research work. However, global stabilization of a rigid body attitude is subject to topological constraints implicit in the attitude state space $\mathrm{SO(3)}$ (i.e., there exists at least one other singularity besides a desired attracting equilibrium \cite{Koditschek1988,Maithripala2006}). For example, the strongest convergence behavior that a continuous state-feedback controller (e.g., \cite{Bullo1999,Du2011,Akhtar2021,Dongare2021}) can attain is \emph{almost global} asymptotic stability, where the region of attraction excludes a set of zero measure \cite{Chaturvedi2011}. Moreover, it is impossible to apply (even discontinuous) state-feedback controller to robustly and globally stabilize a rigid body attitude \cite{Sanfelice2006}. The topological issues appear more conspicuous in the parametrization of the rotation matrix. In fact, there exists no 3-component representations of $\mathrm{SO(3)}$ that can be both global and non-singular \cite{Stuelpnagel1964}. Although unit quaternions can globally represent $\mathrm{SO(3)}$ without singularities, the quaternion-based control law may not be globally well-defined on $\mathrm{SO(3)}$ because unit quaternions double-covers $\mathrm{SO(3)}$ \cite{Bhat2000}.

Recently, the synergistic control has been employed to achieve robust and global attitude tracking under the framework of the hybrid dynamical systems presented in \cite{Goebel2012,Sanfelice2021}. Its key ingredient is the synergistic potential functions that can be classified into central \cite{Mayhew2011b,Casau2015,Berkane2017a,Berkane2017b,Wang2021} and non-central \cite{Mayhew2013,Lee2015} based on whether the common critical points of the potential functions on $\mathrm{SO(3)}$ are mapped to the desired set. A hybrid switching mechanism is designed to coordinate the state-feedback laws induced from the potential functions, so as to avoid the unwanted critical points and thus to ensure the global stability \cite{Sanfelice2021}. The main difference between the central and non-central synergistic designs lies in that each individual state-feedback law in the former can (non-globally) asymptotically stabilize the plant to the desired set; however, some state-feedback laws in the latter are designed to expel the existing unwanted critical points but destabilize the desired set. From a practical standpoint, the central design is more robust than the non-central in consideration of the possible fault of the switching mechanism. On the other hand, the non-central synergistic potential functions on the unit quaternions has been proposed in \cite{Mayhew2011,Gui2016,Schlanbusch2012,Schlanbusch2015,Hashemi2021,Gui2021,Naldi2017}, where two state-feedbacks are implemented to stabilize the two desired points respectively.

One of the most popular candidates for constructing the synergistic family on $\mathrm{SO(3)}$ is the \emph{modified trace function}, because it can be naturally constructed by using the weighted inertial vectors that are usually available for recovering the rigid-body attitude in practice. Particularly, the state-feedback law induced from the modified trace function can obviate the attitude reconstruction through an intricate design \cite{Tayebi2013}. The earliest design for synergistic family is to smoothly perturb a modified trace function via \emph{angular warping} along two opposite directions, although only providing a numerical procedure for the construction \cite{Mayhew2011b}. The term ``angular warping'' means perturbing the error attitude about different directions by the error-dependent angle. Further, the explicit solution of the warping directions for the specific modified trace functions was proposed in \cite{Casau2015}, yet the lower bound of the synergistic gap remains undetermined. On the other hand, \cite{Mayhew2013} put together a non-centrally synergistic family through translating, scaling, and biasing modified trace functions. Although a sufficient condition was also given to guarantee the synergy property, the appropriate parameters have to be tuned carefully following the guideline. Recently, \cite{Berkane2017a} devised a centrally synergistic family based on specific modified trace function and provided the explicit expression of the lower bound of the synergistic gap. Subsequently, one of the unresolved cases was addressed in \cite{Berkane2017b}. On the other hand, \cite{Lee2015} created a non-centrally synergistic family through a direct comparison between the directions of the current and desired attitudes. In addition, \cite{Wang2021} put forward a new hybrid synergistic control strategy where the auxiliary variable for the switching mechanism is defined on a compact and connect space instead of an index set. A more general discussion about the synergistic control for globally asymptotically tracking on compact smooth manifolds was presented in \cite{Casau2020}. 

To the best of our knowledge, the problem of generating a centrally synergistic family from single modified trace function has not been totally resolved using angular warping. The main difficulty lies in ensuring the following conditions hold at the same time: 1) the separation of the unwanted critical points of the potential functions in the family; 2) the existence of a lower evaluation in the family at the unwanted critical points. It becomes more challenging when there exists an infinite set of unwanted critical points excluding the maximum points. This case was thus not discussed in \cite{Mayhew2011b,Mayhew2013,Berkane2017b,Berkane2017a,Casau2015,Wang2021}.

The main contributions of this paper are as follows. First, a set of the warping directions are designed such that a centrally synergistic family can be generated from any single modified trace function via angular warping. Our method can cope with the unresolved case in the literature and thus is generally applicable to global attitude tracking using the inertia vectors measurements. Second, the switching condition is refined for the synergistic control such that only a subset of the potential functions in the family needs to be evaluated at each update. This can reduce the computation cost in practice, because the traditional synergistic control requires evaluating all the potential functions in the family. Furthermore, by virtue of the warping angle function in \cite{Berkane2017a}, the positive lower bound of the synergistic gap at the unwanted critical points is shown explicitly under the refined strategy. Finally, the proposed synergistic functions are implemented in the refined synergistic control to achieve robust and global attitude tracking.

The rest of this paper is organized as follows. We present some preliminaries on synergistic control and formalize the problem in Section~\ref{sec:pre}. In Section~\ref{sec:main}, we introduce the approach to the construction of the centrally synergistic family and its application to global attitude tracking. Illustrative simulation results are given in Section \ref{sec:sim}. Conclusions and perspectives are given in Section \ref{sec:con}. The proofs of main results are deferred to the Appendix.

\section{Preliminaries and Problem Statement} \label{sec:pre}

\subsection{Notation}
In this paper, we denote by $\mathbb{R}_{\geq 0}$, $\mathbb{Z}$, and $\mathbb{N}$, the set of nonnegative real numbers, integers, and nonnegative integers, respectively. The standard Euclidean norm is given by $|x| \coloneqq \sqrt{x^\top  x}$ for each $x \in \mathbb{R}^n$. Given matrices $A,B \in \mathbb{R}^{m \times n}$, their Frobenius inner product is defined as $\langle A , B \rangle = \operatorname{tr} (A^\top B)$. The unit $n$-sphere is defined by $\mathbb{S}^n = \{ x \in \mathbb{R}^{n+1} : |x| = 1 \}$, and the closed $n$-ball with radius $r$ is defined by $\bar{\mathbb{B}}_r^n = \{x \in \mathbb{R}^{n} : |x| \leq r\}$. Let $I$ (or $I_n$) denote the identity matrix (of order $n$). The 3-dimensional rotation group is $\mathrm{SO(3)}= \{R \in \mathbb{R}^{3\times 3}:R^\top R = I, \mathrm{det}(R) = 1 \}$, and the space of 3-by-3 skew-symmetric matrices is $\mathfrak{so}(3) = \{\Omega \in \mathbb{R}^{3\times 3}: \Omega^\top  = - \Omega   \}$. The skew-symmetric operator $(\cdot)^\wedge: \mathbb{R}^3 \to \mathfrak{so}(3)$ is defined as  $x^\wedge = 
\begin{bsmallmatrix}
    0 & -x_3 & x_2 \\
    x_3 & 0 & -x_1 \\
    -x_2 & x_1 & 0
\end{bsmallmatrix}$.
One can verify that $x^\wedge y = x \times y$ for all $x,y \in \mathbb{R}^3$, where $\times$ denotes the cross product in $\mathbb{R}^3$. The inverse of $(\cdot)^\wedge$ is denoted by $(\cdot)^\vee$ such that $ (x^\wedge)^\vee = x $. The antisymmetric projection is defined as $\mathbb{P}_a : \mathbb{R}^{3\times 3} \to \mathfrak{so}(3)$ such that $\mathbb{P}_a (X) = (X - X^\top )/2$. Let us define a composition map $\psi : \mathbb{R}^{3 \times 3} \to \mathbb{R}^3$ such that $\psi (X) = (\mathbb{P}_a (X))^\vee$. For each $A = A^\top \in \mathbb{R}^{n \times n}$, we denote the set of all unit-eigenvectors of $A$ by $\mathcal{E}(A)$. The $i$th element of $\mathcal{E}(A)$ and the associated eigenvalue are denoted by $(\lambda_i^A,v_i^A) $. The maximum and minimum eigenvalues of $A$ are denoted by $\lambda_{\max}^A$ and $\lambda_{\min}^A$, respectively.

Given a finite subset $\mathcal{Q} \subset \mathbb{N}$, we denote by $\mathcal{C}^1 (\mathrm{SO(3)} \times \mathcal{Q},\mathbb{R})$ the set of functions $V : \mathrm{SO(3)} \times \mathcal{Q} \to \mathbb{R}$ such that the map $X \mapsto V(X,q) $ is continuously differentiable for each $q \in \mathcal{Q}$. Given a function $V \in \mathcal{C}^1 (\mathrm{SO(3)} \times \mathcal{Q},\mathbb{R})$, we denote by $\nabla V (X,q) $ the gradient of $V$ relative to $X$ with $q$ considered to be constant, and by $\operatorname{Crit} (V) = \{(X,q) \in \mathrm{SO(3)} \times \mathcal{Q} : \rho_V (X,q) = 0\}$ the set of its critical points, where the function $\rho_V : \mathrm{SO(3)} \times \mathcal{Q} \to \mathbb{R}^3 $ is defined as $\rho_V (X,q) = \psi(X^\top  \nabla V(X,q)) $ \cite{Mayhew2013}. Note that $\operatorname{Crit} (V)$ are the points where $V$ has no infinitesimal change along the trajectories $\dot{X}  = X \omega^\wedge, \omega \in \mathbb{R}^3 $. The rotation matrix is often parametrized in terms of a rotation angle $\theta \in \mathbb{R}$ and an axis $u \in \mathbb{S}^2$ by the Rodrigues formula $\mathcal{R}_a(\theta, u) = I + u^\wedge \sin\theta + ( u^\wedge )^2 (1-\cos \theta)$.

\subsection{Hybrid dynamical systems}

A set-valued mapping $E$ from $\mathbb{R}^m$ to $\mathbb{R}^n$ associates every point $x \in \mathbb{R}^m$ with a subset of $\mathbb{R}^n$ and is denoted by $E : \mathbb{R}^m \rightrightarrows \mathbb{R}^n$. A hybrid system $\mathcal{H}$ defined on $\mathbb{R}^n$ has the data $(\mathcal{F}, F, \mathcal{J}, G)$ and is given by 
\begin{equation*}
  \mathcal{H} :
  \begin{cases}
    \dot{x} \in F(x) & x \in \mathcal{F},\\
    x^+ \in G(x) & x\in \mathcal{J},
  \end{cases}
\end{equation*}
where $x \in \mathbb{R}^n$ is the state, $F: \mathbb{R}^n \rightrightarrows \mathbb{R}^n$ is the \emph{flow map} capturing the continuous dynamics on the \emph{flow set} $\mathcal{F} \subset \mathbb{R}^n$, $G: \mathbb{R}^n \rightrightarrows \mathbb{R}^n$ is the \emph{jump map} capturing the discontinuous dynamics on the \emph{jump set} $\mathcal{J} \subset \mathbb{R}^n$, and $x^+$ indicates the values of the state after the jump. A solution $x$ to $\mathcal{H}$ is parametrized by $(t,j) \in \mathbb{R}_{\geq 0} \times \mathbb{N}$, where $t $ denotes the ordinary time and $j$ denotes the jump time. The notions of solutions to a hybrid system, its hybrid time domain, type of solutions (maximal and complete solutions), asymptotic stability, and invariance are referred to \cite{Goebel2009,Goebel2012,Sanfelice2021}.

\subsection{Synergistic potential functions on $\mathrm{SO(3)}$}
\begin{defn}\label{def:pfunc}
  A continuously differentiable function $V: \mathrm{SO(3)} \to \mathbb{R}$ is the \emph{potential function} relative to $\{I\} $ if $V(I) = 0$ and $V(X) > 0$ for all $X \in \mathrm{SO(3)} \setminus \{I\}$. 
\end{defn}
Given a matrix $M  = M^\top \in \mathbb{R}^{3\times 3}$ such that $G \coloneqq \operatorname{tr} (M) I -M $ is symmetric positive definite, we define $\Psi_M: \mathrm{SO(3)} \to \mathbb{R} $ as
\begin{equation}\label{eq:modifiedPF}
    \Psi_M(X) =  \operatorname{tr} (M(I - X)) . 
\end{equation} 
The function $\Psi_M$ is called \emph{modified trace function} and has been widely used in attitude control \cite{Mayhew2013,Casau2015,Berkane2017a,Wang2021}. The next lemma states some useful properties of $\Psi_M$.
  
\begin{lem}[{\cite[Lemma 2]{Berkane2017a},\cite[Proposition 11.31]{Bullo2004}}] \label{lem:property_MPF}
    The following properties hold for $\Psi_M$ defined as in \eqref{eq:modifiedPF}.
    \begin{enumerate}[1)]
        \item The eigenvalues and eigenvectors of $M$ and $G$ satisfy
        \begin{align}
          \lambda_i^G &= \operatorname{tr} (M) - \lambda_i^M, & v_i^M &= v_i^G, & i = \{1,2,3\}. \label{eq:eigGM}
        \end{align} 
        \item The gradient and critical points of $\Psi_M$ are given by
        \begin{align}
          \nabla \Psi_M (X) & = X \mathbb{P}_a (MX) , \label{eq:gradMPF}\\
          \operatorname{Crit} (\Psi_M) &= \{I\} \times \mathcal{R}_a (\pi, \mathcal{E}(M)). \label{eq:critMPF}
        \end{align}
        \item For each $v \in \mathcal{E}(G)$ and each $(\theta , u) \in \mathbb{R} \times \mathbb{S}^2$, one has
        \begin{align}
            \Psi_M (\mathcal{R}_a(\theta,u)) &= (1-\cos(\theta)) u^\top  G u , \label{eq:EvaModPF} \\
            \Psi_M (\mathcal{R}_a(\pi,v) \mathcal{R}_a(\theta,u) ) &= 2 \lambda^G -  (1-\cos(\theta)) \Delta (v,u), \label{eq:EvaModPFPi}
        \end{align}
        where $\lambda^G$ is the eigenvalue associated to $v$ and $\Delta (v,u)$ is defined as follows.
        \begin{itemize}
            \item If $M$ has one distinct eigenvalue $\lambda_i^M = \lambda > 0,\ i = 1,2,3 $, then $\mathcal{E}(M) = \mathbb{S}^2$ and $\Delta(v,u) = \lambda^G (v^\top  u)^2 $. 
            \item If $M$ has two distinct eigenvalues $\lambda_1^M = \lambda^M_2 \neq \lambda^M_3$, then $\mathcal{E}(M) =\{  v_3 \} \cup \{v_{12} = v_1 \cos (t) + v_2 \sin(t): t \in \mathbb{R} \}  $, and $\Delta(v_{3},u) = \lambda^G_3 - \lambda^G_2(1-(u^\top  v_3)^2)$ and $\Delta(v_{12},u) = (1-(u^\top  v_3)^2) (\lambda^G_2 - \lambda^G_3 \sin^2 (\phi))$, where $\phi = \angle (v_{12},u^\perp)$ and $u^\perp \coloneqq (I - v_3 v_3^\top )u $ is the projection of $u$ on the plane $\mathrm{span}\{v_1,v_2\}$.
            \item If $M$ has distinct eigenvalues $ \lambda^M_1 < \lambda^M_2 < \lambda^M_3$, then $\mathcal{E}(M) = \{v_1,v_2,v_3\}$ and $\Delta(v_i,u) = \lambda^G_i - (u^\top v_j)^2 \lambda^G_k - (u^\top  v_k)^2 \lambda^G_j$ for $i \neq j \neq k $. 
        \end{itemize}
    \end{enumerate}
\end{lem}

\begin{figure}[t]
  \centering
  \includegraphics[width = 8cm]{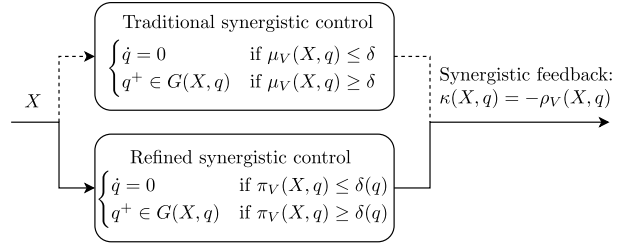}
  \caption{Logic implemented by synergistic feedback control. $\dot{q} = 0$ represents that the switch is not triggered; $q^+$ denotes the new value of $q$ when the switch is triggered; $G(X,q) = \left\{p \in \mathcal{Q}: {\arg\min}_{p \in  \mathcal{Q}} V(X,p) \right\}$.}
  \label{fig:flowchart}
\end{figure}

Now, we give the definition of the centrally synergistic potential functions on $\mathrm{SO(3)}$.
\begin{defn}[\cite{Mayhew2013,Berkane2017a}] \label{def:cenSynFunc}
  Let $\mathcal{Q} \subset \mathbb{N}$ be a nonempty finite set and $V \in \mathcal{C}^1 (\mathrm{SO(3)} \times \mathcal{Q},\mathbb{R})$ such that for each $q\in \mathcal{Q}$, the map $X \mapsto V(X,q)$ is a potential function. Define the set 
  \begin{equation}
    \mathcal{A} \coloneqq \{I \} \times \mathcal{Q}. \label{eq:setpoint_defn}
  \end{equation}
  The \emph{synergistic gap} of $V$ is the function $\mu_V \in \mathcal{C}^1 (\mathrm{SO(3)} \times \mathcal{Q},\mathbb{R})$ defined as $\mu_V(X,q) = V(X,q) - \min_{p\in \mathcal{Q}} V(X,p)$. Let $\delta > 0$. The family $V $ is called \emph{centrally synergistic} relative to $\mathcal{A} $ with \emph{gap exceeding} $\delta$ if 
  \begin{align} \label{eq:cenSynergy}
    \mu_V(X,q) > \delta && \forall (X,q) \in  \operatorname{Crit} (V) \setminus \mathcal{A}.
  \end{align} 
\end{defn}
\begin{rem}
    It is noteworthy that the potential functions $X \mapsto V(X,q)$ for each $q \in \mathcal{Q}$ have at least three unwanted critical points by Morse theory \cite{Koditschek1988,Maithripala2006}. The condition of \eqref{eq:cenSynergy} implies that these potential functions have separate unwanted critical points and only one common critical point, i.e., the identity. Furthermore, if $V(X,q)$ is taken as Lyapunov function, when $V(X,q)$ stops decreasing along trajectories of $X$ at the unwanted critical points by the $q$-th state-feedback law, \eqref{eq:cenSynergy} ensures that the state of the plant can be pushed away by another state-feedback law indexed by ${\arg\min}_{p\in \mathcal{Q}} V (X,p)$; see the traditional synergistic control depicted in Figure~\ref{fig:flowchart}.
\end{rem} 
The angular warping presented in \cite{Mayhew2011b,Casau2015,Berkane2017a} is an effective tool to extend a modified trace function to a synergistic family, which is defined by $\mathcal{T}: \mathrm{SO(3)} \times \mathcal{Q} \to \mathrm{SO(3)} $ as
\begin{equation} \label{eq:angularTrans}
    \mathcal{T}(X,q) = X \mathcal{R}_a (\theta(X), u_q),
\end{equation}
where $\theta: \mathrm{SO(3)} \to \mathbb{R}$ is a continuously differentiable function and $u_q \in \mathbb{S}^2$ is the warping direction. The next Lemma \ref{lem:synergyCon} provides the condition on the local diffeomorphism, so as to compute the new critical points through evaluating the inverse of $\mathcal{T}$ at the critical points of the original potential function.
 
\begin{lem}[{\cite[Lemma 1]{Berkane2017a}}] \label{lem:synergyCon}
  Consider the map $\mathcal{T}$ defined as \eqref{eq:angularTrans}. The following statements hold.
  \begin{enumerate}[1)]
    \item For a given $q \in \mathcal{Q}$ and trajectories of $\dot{X} = X \omega^\wedge$, the time derivative of $\mathcal{T}(X,q)$ is $\dot{\mathcal{T}}(X,q) = \mathcal{T}(X,q) (\Theta(X,q) \omega )^\wedge$ with the function $\Theta : \mathrm{SO(3)} \times \mathcal{Q} \to \mathbb{R}^{3 \times 3}$ defined as $\Theta(X,q) \coloneqq \mathcal{R}_a(\theta(X),u_q)^\top  + 2u_q \psi(X^\top  \nabla \theta (X))^\top $.  
    \item Given the potential function $\Psi_M$ by \eqref{eq:modifiedPF}, we define the function $V : \mathrm{SO(3)} \times \mathcal{Q} \to \mathbb{R}$ as $V(X,q) = \Psi_M (\mathcal{T}(X,q))$. If $\det (\Theta(X,q)) \neq 0$ for all $ (X,q) \in \mathrm{SO(3)} \times \mathcal{Q}$ and $\mathcal{T}^{-1} (I) = \mathcal{A}$ with $\mathcal{A}$ given by \eqref{eq:setpoint_defn}, then $V$ is positive definite relative to $\mathcal{A}$ and $\operatorname{Crit} (V) = \mathcal{T}^{-1}(\operatorname{Crit} (\Psi))$. 
  \end{enumerate}
\end{lem}

\subsection{Problem statement}
Let us denote by $ R \in \mathrm{SO(3)}$ the orientation of the body-fixed frame relative to the inertial reference frame, and by $R_d \in \mathrm{SO(3)}$ the desired reference attitude. We define the left attitude error as $\tilde{R} = R R_d^\top  $ \cite{Bullo1999}. Since there exists no single sensor that can measure the attitude $R$ directly, the inertial vector measurements are commonly used as the indirect observation. Therefore, we will make the following standard assumptions.
\begin{assum}[{\cite{Mayhew2011,Bullo1999,Gui2016}}]\label{ass:vector}
  There exist $n \geq 2$ known unit vectors, denoted by $a_i$ for $i=1,...,n,$ in the inertial reference frame being measured in the body-fixed frame as $b_i = R^\top  a_i$. At least two of $a_i$ are noncollinear. Additionally, there exists the measurement of the angular velocity $\omega$.
\end{assum}
\begin{rem}
  Assumption \ref{ass:vector} is a standard assumption in attitude tracking problem that imposes the mildest condition on the attitude reconstruction from the vector measurements; see \cite{Tayebi2013,Berkane2017a}. The attitude tracking under Assumption \ref{ass:vector} is referred to as the full-state measurement tracking.
\end{rem} 
The attitude tracking is formulated as the minimization of the cost function $J_0 : \mathrm{SO(3)} \times \mathrm{SO(3)} \to \mathbb{R}_{\geq 0}$ defined as 
\begin{equation} \label{eq:wahba}
  J_0 (R,R_d) = \frac{1}{2} \sum_{i=1}^{n} w_i |b_i - R_d^\top  a_i |^2 = \operatorname{tr} (M_a (I - \tilde{R}))
\end{equation}
where $w_i > 0$ is the positive weight that represents the confidence of the $i$-th vector measurement \cite{Chang2017}, and $M_a = \sum_{i} w_i a_i a_i^\top $. Therefore, the function $J_0$ can be treated as the potential function with respect to $\tilde{R}$, which implies that $J_0 (R,R_d) = 0$ if and only if $R = R_d$. By Assumption \ref{ass:vector}, $M_a$ has two possible cases: 1) $M_a$ is positive definite; 2) $M_a$ is positive semi-definite with one eigenvalue zero.

To achieve global attitude tracking, an intuitive approach is to construct a family of synergistic function from the modified trace function with given $M = M_a$. This has been partly solved in \cite{Mayhew2013,Casau2015,Berkane2017b,Berkane2017a}, but some typical cases remain unresolved, e.g., the scenario where two vector measurements or three orthogonal measurements with the equivalent weights exist.

In this work, our goal is to construct the centrally synergistic functions through angular warping from the modified trace function given by \eqref{eq:wahba} with all possible $M_a$.

\section{Main Result} \label{sec:main}
\subsection{Construction of synergistic potential functions}
For a synergistic family satisfying Definition \ref{def:cenSynFunc}, the synergistic control needs to compute the synergistic gap at each update, and thus all the potential functions in the family have to be evaluated. The next lemma states that the number of the potential functions to be evaluated at each update can be reduced under some conditions.  
\begin{lem}\label{lem:cenSynergyLocal}
    Let $\mathcal{Q} \subset \mathbb{N}$ be a nonempty finite set and the function $V \in \mathcal{C}^1 (\mathrm{SO(3)} \times \mathcal{Q},\mathbb{R})$ with the map $X \mapsto V(X,q)$ being a potential function for each $q \in \mathcal{Q}$. Then, the following statements hold.
    \begin{enumerate}[1)]
        \item $\mathcal{A} \subset \operatorname{Crit} (V)$ with $\mathcal{A} $ given by \eqref{eq:setpoint_defn}.
        \item $V$ is centrally synergistic relative to $\mathcal{A}$ if and only if there exists a subset $\mathcal{Q}_q \subseteq \mathcal{Q}$ for each $q \in \mathcal{Q}$ and any function $\delta: \mathcal{Q} \to \mathbb{R}_{\geq 0}$ such that 
        \begin{align} \label{eq:cenSynergyLocal}
            \pi_V(X,q) &> \delta(q) > 0 & \forall (X,q) \in  \operatorname{Crit} (V) \setminus \mathcal{A} ,
        \end{align}
        where $\pi_V \in \mathcal{C}^1 (\mathrm{SO(3)} \times \mathcal{Q},\mathbb{R})$ is called \emph{refined synergistic gap} defined as 
        \begin{equation}
          \pi_V(X,q) = V(X,q) - \min_{p\in \mathcal{Q}_q} V(X,p). \label{eq:refined_Syn_gap}
        \end{equation}
    \end{enumerate}
\end{lem}

\begin{rem} 
    Invoking Lemma~\ref{lem:cenSynergyLocal}, the switching condition in the synergistic control can be rephrased by using the refined synergistic gap as shown in Figure~\ref{fig:flowchart}. Moreover, the number of the potential functions to be evaluated at each update can be reduced compared to the traditional synergistic control. 
\end{rem}
\begin{rem}
  The refined synergistic gap is a conservative substitute for the synergistic gap as $\mu_V(X,q) \geq \pi_V (X,q)$ always holds. Additionally, the synergistic condition \eqref{eq:cenSynergy} is equivalent to the refined condition \eqref{eq:cenSynergyLocal}, provided that $\mathcal{Q}_q  \equiv \mathcal{Q}$ or the cardinality of $\mathcal{Q} $ is two.
\end{rem}

In what follows, we demonstrate that it is possible to attain a centrally synergistic family from a single modified trace function \eqref{eq:modifiedPF} through the angular warping \eqref{eq:angularTrans} with the collection of the warping directions denoted by $\mathcal{U} = \cup_{q \in \mathcal{Q}} \{u_q \}$. Furthermore, let $\mathcal{U}_q = \cup_{p \in \mathcal{Q}_q} \{u_p\} \subseteq \mathcal{U} $ for each $q \in \mathcal{Q} $ be the collection of the warping directions associated to $\mathcal{Q}_q$ that is defined in Lemma~\ref{lem:cenSynergyLocal}. With these notations, the next lemma states the necessary condition for the synergy property of \eqref{eq:cenSynergyLocal}.

\begin{lem} \label{lem:condSynergMPF}
  Let $\mathcal{Q} \subset \mathbb{N}$ be a nonempty finite set. Consider the angular warping $\mathcal{T}$ in \eqref{eq:angularTrans} with the warping angle $\theta(X) : \mathrm{SO(3)} \to [0,\pi)$ being a potential function. Assume that $\mathcal{T}^{-1} (I) = \mathcal{A}$ and that $\det (\Theta(X,q)) \neq 0$ for all $(X,q) \in \mathrm{SO(3)} \times \mathcal{Q}$. The functions \eqref{eq:modifiedPF} and \eqref{eq:angularTrans} are composed to yield the family 
  \begin{equation} \label{eq:warpMPF}
      V (X,q) = \Psi_M (\mathcal{T}(X,q) ). 
  \end{equation} 
  Then, the following statements hold.
  \begin{enumerate}[1)]
    \item The set of the unwanted critical points of $V$ is given by 
    \begin{equation}
      \operatorname{Crit} (V) \setminus \mathcal{A} = \cup_{v\in \mathcal{E}(M)} \mathcal{T}^{-1} (\mathcal{R}_a(\pi,v)).
    \end{equation}
    \item Let $(Y,q) \in \operatorname{Crit} (V) \setminus \mathcal{A}  $ and $\mathcal{R}_a(\theta_{pq},u_{pq})\coloneqq \mathcal{R}_a(\theta(Y),-u_q) \mathcal{R}_a(\theta (Y),u_p)$ with $(\theta_{pq}, u_{pq}) \in R \times \mathbb{S}^2$ and $u_p \neq u_q \in \mathcal{U}$. Then $\mathcal{R}_a(\theta_{pq}, u_{pq})\neq I$ and the rotation axis $u_{pq}$ is well-defined
    \begin{subequations}
      \begin{align}
        &u_{pq} = \frac{ ( u_p - u_q ) \frac{\sin(\theta(Y))}{2} + u_p  \times u_q \sin^2 (\frac{\theta(Y)}{2}) }{\sin(\frac{\theta_{pq}}{2})}, \label{eq:u_pq_def} \\
        &\cos(\frac{\theta_{pq}}{2}) = \cos^2 (\frac{\theta(Y)}{2}) + u_p^\top  u_q\sin^2 (\frac{\theta(Y)}{2}). \label{eq:theta_pq_def}
      \end{align}
    \end{subequations}
    \item If $V$ is centrally synergistic relative to $\mathcal{A}$, there exists a subset $\mathcal{Q}_q \subseteq \mathcal{Q}$ for each $q \in \mathcal{Q}$ such that 
    \begin{equation} \label{eq:cond_synerg}
      \max_{p \in \mathcal{Q}_q \setminus \{q\}}\Delta (v,u_{pq}) > 0 
    \end{equation}
    for all $v \in \mathcal{E}(M)$ satisfying $(Y,q) \in \mathcal{T}^{-1 } (\mathcal{R}_a(\pi, v))$, where the function $\Delta$ is given by \eqref{eq:EvaModPFPi}. 
  \end{enumerate}
\end{lem}
\begin{rem}
    The assumption on the range of the warping angle $\theta$ being $[0,\pi)$ ensures that the angular warping along the opposite direction induces a valid rotation (not equal to $I$), so as to avoid returning to the original unwanted critical point. This is a necessary condition for separating unwanted critical points of $V(X,q)$. 
\end{rem} 
\begin{rem}
  The condition of \eqref{eq:cond_synerg} is the necessary condition for the synergy property of \eqref{eq:cenSynergyLocal}. The mildest condition \eqref{eq:cond_synerg} occurs when $\mathcal{Q}_q = \mathcal{Q}$ holds for all $q \in \mathcal{Q}$.
\end{rem}
The next theorem gives the set of warping directions $\mathcal{U}$ and its subset $\mathcal{U}_q$ for each $q \in \mathcal{Q}$ to guarantee the necessary condition of the central synergy given by \eqref{eq:cond_synerg}.  

\begin{thm}\label{thm:SelectDirctions}
  Let all assumptions and definitions of Lemma~\ref{lem:condSynergMPF} hold. We write $(v_1,v_2,v_3)$ for the orthonormal eigenbasis of $M$ where $v_i$ is associated to the eigenvalue $\lambda_i^M$. Then, \eqref{eq:cond_synerg} satisfies if the set of warping directions $\mathcal{U}$ and its subsets $\mathcal{U}_q$ for each $q \in \mathcal{Q}$ are selected as follows. 
  \begin{enumerate}[1)]
    \item If $\lambda_i^M = \lambda >0$ for $i  = 1,2,3 $, $\mathcal{U} = \{\pm v_1,\pm v_2,\pm v_3\}$. For each $q \in \mathcal{Q}$, $\mathcal{U}_q = \{u_p \in \mathcal{U}: u_p^\top  u_q = 0\}$.
    \item If $\lambda_1^M = \lambda^M_2 > \lambda^M_3 > 0$, $\mathcal{U}  = \{\pm v_1,\pm v_2\}$. For each $q \in \mathcal{Q}$, $\mathcal{U}_q = \{u_p \in \mathcal{U} : u_p^\top  u_q =  0\}$. 
    \item If $\lambda_1^M = \lambda^M_2 > \lambda^M_3 \geq 0$, $\mathcal{U}  = \{v_1 \cos (\frac{n\pi}{3}) + v_2 \sin(\frac{n\pi}{3}) : n \in \mathbb{Z}\}$. For each $q \in \mathcal{Q}$, $\mathcal{U}_q = \{-u_q\} \cup \{ u_p \in \mathcal{U}: u_p^\top  u_q = \frac{1}{2}\}$.
    \item If $0 < \lambda_1^M = \lambda^M_2 < \lambda^M_3 $, $\mathcal{U} = \{u,-u\} \subset \mathbb{S}^2$ such that $0 < 1 - (u^\top  v_3)^2 < (\lambda_3^G / \lambda_2^G)$.
    \item If $M$ has three distinct eigenvalues ($0 \leq \lambda^M_1 < \lambda^M_2 < \lambda^M_3$), $\mathcal{U} = \{u,-u\} \subset \mathbb{S}^2$ such that $\Delta(v_2,u) > 0$ and $\Delta(v_3,u)> 0$.
  \end{enumerate}
\end{thm}

\begin{rem}
    Theorem \ref{thm:SelectDirctions} provides a complete solution to guarantee the potential functions of \eqref{eq:warpMPF} are centrally synergistic relative to $\mathcal{A}$ for all $M \geq 0$ with $\mathrm{rank}(M) \geq 2$. Moreover, the items (1-3) figure out the cases that are difficult and unresolved in \cite{Mayhew2013, Berkane2017a}. As pointed in \cite{Berkane2017a}, it is impossible to accomplish the task for the cases in the items (1), (2), and (3) by using only two warping directions designed. The main difficulty is to guarantee the condition \eqref{eq:cond_synerg} for all the unwanted critical points. Moreover, although the similar choice of the warping directions for the item (1) was proposed in \cite{Casau2015,Berkane2017b}, we will explicitly express the lower bound of the refined synergistic gap in Theorem \ref{thm:explicit_syngap}. The optimal choice of warping directions for the items (4) and (5) was detailed in \cite{Berkane2017a}, so omitted here.  
\end{rem} 
\begin{rem}
  Theorem \ref{thm:SelectDirctions} specifies the $\mathcal{U}_q$ for each $q \in \mathcal{Q}$. In consequence, the refined synergistic control can be applied as shown in Figure~\ref{fig:flowchart} and so the number of the potential functions to be evaluated at each update, denoted by $N$, is less compared to the traditional synergistic control. Specifically, $N$ is reduced from $6$ to $3$ for the item (1), from $4$ to $3$ for the item (2), and from $6$ to $3$ for the item (3). For the item (4) and (5), $N$ does not change since the synergistic family is composed of two functions.
\end{rem} 
The remaining work for the implementation of Theorem \ref{thm:SelectDirctions} is to choose the eligible warping angle function $\theta$ and then to determine the positive lower bound of the synergistic gap. The following proposition allows one to compute the unwanted critical points explicitly, so as to compute the synergistic gap.

\begin{prop}[{\cite[Lemma 3]{Berkane2017a}}] \label{prop:warpAngle}
    Consider the family of functions \eqref{eq:warpMPF}. If the angular warping angle is defined as 
    \begin{equation} \label{eq:WarpAngle_MPF}
        \theta (X) = 2\arcsin \left(  {k}{(2\lambda_{\max}^G)^{-1}} \Psi_M(X) \right) ,
    \end{equation}
    where $0 <k  < 1/ \sqrt{6 - \max \{1,4\xi^2\}} $ is a positive gain and $\xi = \lambda_{\min}^G / \lambda_{\max}^G$. Then, $\mathcal{T}^{-1} (I) = \mathcal{A}$ and $\det (\Theta(X,q)) \neq 0$, $\forall (X,q) \in \mathrm{SO(3)} \times \mathcal{Q}$. 
\end{prop}

By Lemma \ref{lem:cenSynergyLocal}, we denote by $\bar{\delta}_q$ the lower bound of the refined gap at the undesired critical points of $V(X,q)$ such that 
\begin{equation} \label{eq:bar_delta_q}
  \bar{\delta}_q = \min \left\{\pi_V (X,p): (X,p)\in \operatorname{Crit} (V) \setminus \mathcal{A}, p = q \right\}.
\end{equation} 
With the knowledge of $\bar{\delta}_q$, it is convenient to select the hysteresis gap $\delta(q) < \bar{\delta}_q$ used in \eqref{eq:cenSynergyLocal}. The next theorem gives the explicit expression of $\bar{\delta}_q$ of the family of \eqref{eq:warpMPF} together with \eqref{eq:WarpAngle_MPF}. 

\begin{thm} \label{thm:explicit_syngap}
  Let all assumptions and definitions of Theorem~\ref{thm:SelectDirctions} hold. Consider the family of functions \eqref{eq:warpMPF} with the warping angle in \eqref{eq:WarpAngle_MPF}, and warping directions given by Theorem \ref{thm:SelectDirctions}. Then, $\bar{\delta}_q$ defined as in \eqref{eq:bar_delta_q} is determined as follows. 
  \begin{enumerate}[1)]
    \item If $\lambda_i^M = \lambda >0$ for $i  = 1,2,3 $, then for every $q \in \mathcal{Q}$, 
    \begin{align} 
      \bar{\delta}_q & = 2\lambda  \min \bigl\{  k^2, 2\Xi_1^2 (1 - \Xi_1^2) \bigr\} ,  \label{eq:refined_sy_gap_1eigx}
    \end{align}
    where $\Xi_1 \coloneqq 2k (1 + \sqrt{1 + 4k^2})^{-1} $. 
    \item  If $\lambda_1^M = \lambda^M_2 > \lambda^M_3 > 0$, then for every $q \in \mathcal{Q}$,
    \begin{equation} \label{eq:refined_sy_gap_2eigx}
      \begin{split}
        \bar{\delta}_q =2  \lambda_3^G \min \bigl\{ &  \Xi_{21}^2 (1 + (1-2\xi)(1-\Xi_{21}^2) ), \\ 
        &  \Xi_{22}^2 (1 - \Xi_{22}^2) (2\xi - 1) \bigr\}  ,
      \end{split}
    \end{equation}
    where $\Xi_{21} \coloneqq 2k (1 + \sqrt{1 + 4k^2 (1-\xi)})^{-1} $ and $\Xi_{22} \coloneqq 2k\xi (1 + \sqrt{1 + 4k^2 \xi^2})^{-1}$. 
    \item If $\lambda_1^M = \lambda^M_2 > \lambda^M_3 \geq 0$, then for every $q \in \mathcal{Q}$,
    \begin{equation} \label{eq:refined_sy_gap_2eigx1}
        \begin{split}
          \bar{\delta}_q \geq & \min \lambda_3^G \biggl\{   \max \Bigl\{  \frac{1}{2}  \Xi_{21}^2 (3 + (1-4\xi)(1-\Xi_{21}^2) ) ,\\
          &   8  \Xi_{21}^2 (1 - \Xi_{21}^2) (1 - \xi) \Bigr\} ,   2\Xi_{22}^2 (1 - \Xi_{22}^2) (\xi - \frac{1}{4})   \biggr\}.
        \end{split}
      \end{equation}
  \end{enumerate}
\end{thm}
\begin{rem}
    Invoking Theorem \ref{thm:SelectDirctions}, Theorem \ref{thm:explicit_syngap} explicitly expresses the lower bound of the refined synergistic gap \eqref{eq:bar_delta_q}. Of note, the items (2) and (3) in Theorem \ref{thm:SelectDirctions} can cope with the same case and even the former requires less number of the potential functions to be evaluated at each update. However, the refined synergistic gap \eqref{eq:refined_sy_gap_2eigx} will converge to zero as $\xi \to \frac{1}{2}$, which decreases the magnitude of the noise that can be tolerated. 
\end{rem}

\begin{figure}[t]
  \centering
  \includegraphics[width = 8cm]{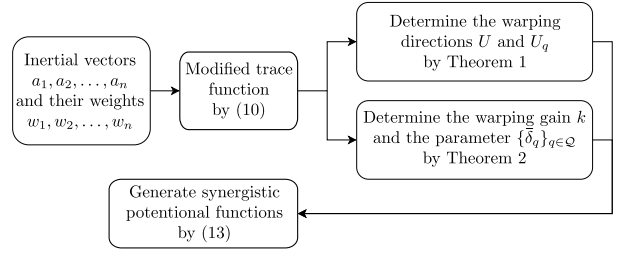}
  \caption{The block diagram of constructing the synergistic potential functions.}
  \label{fig:blk_diag_Construction}
\end{figure}

Figure~\ref{fig:blk_diag_Construction} illustrates the procedure of the proposed approach for constructing the synergistic potential functions. First, it is required to obtain the modified trace function of \eqref{eq:wahba} by using the weighted inertial vectors. Then, one can determine the required parameters by Theorems~\ref{thm:SelectDirctions} and~\ref{thm:explicit_syngap}. Finally, with those parameters, the synergistic potential functions can be given by \eqref{eq:warpMPF}.

\subsection{Application to attitude tracking} 
In this section, we apply the proposed centrally synergistic functions to attain the global attitude tracking. 

The kinematic and dynamic equations of a rigid body are given by \cite{Bullo1999,Mayhew2013}
\begin{equation} \label{eq:body_dyn}
  \begin{matrix}
    \dot{R} = R \omega^\wedge, &
    J \dot{\omega} = - \omega^\wedge J \omega  + \tau, 
  \end{matrix}
\end{equation} 
where $\omega \in \mathbb{R}^3$ is the angular velocity expressed in the body-fixed frame, $J \in \mathbb{R}^{3 \times 3}$ is the constant inertia matrix (positive and symmetric), and $\tau \in \mathbb{R}^3 $ is an external torque to be designed. 

Let $  c_\omega, c_a > 0$ be constant, and in consequence, $\mathcal{W}_d = \mathrm{SO(3)} \times  \bar{\mathbb{B}}^3_{c_\omega}$ be compact. Following \cite{Mayhew2013,Wang2021}, the desired reference trajectory is generated by 
\begin{equation} \label{eq:ref_dyn}
  \begin{rcases}
    \dot{R}_d = R_d \omega_d^\wedge \\
    \dot{\omega}_d \in \bar{\mathbb{B}}^3_{c_a}
  \end{rcases} 
  (R_d  , \omega_d) \in \mathcal{W}_d.
\end{equation}

We define the left velocity error as $\tilde{\omega} = \omega - \omega_d$. Combining \eqref{eq:body_dyn} and \eqref{eq:ref_dyn} yields the error dynamics 
\begin{equation*}
  \begin{matrix}
    \dot{\tilde{R}} = \tilde{R}  (R_d \tilde{\omega})^\wedge,&
    J \dot{\tilde{\omega}} = \Sigma (\tilde{\omega} , \omega_d) \tilde{\omega} - \Phi(\tilde{\omega} ,\omega_d,\dot{\omega}_d) + \tau,
  \end{matrix}
\end{equation*}
where the function $\Sigma: \mathbb{R}^3 \times \mathbb{R}^3 \to \mathfrak{so}(3)$ and $\Phi:\mathbb{R}^3 \times \mathbb{R}^3 \times \mathbb{R}^3 \to \mathbb{R}^3$ are defined as $\Sigma (\tilde{\omega} , \omega_d) =   \bigl(J (\tilde{\omega} + \omega_d)\bigr)^\wedge$ and $\Phi(\tilde{\omega} , \omega_d,\dot{\omega}_d) =  {\omega}_d^\wedge J ( \tilde{\omega} + {\omega}_d ) + J \dot{\omega}_d $, respectively. Following \cite{Mayhew2013,Wang2021}, we define the extended state space and state as $\mathcal{W}_z = \mathrm{SO(3)} \times \mathbb{R}^3 \times \mathcal{W}_d$ and $z = (\tilde{R},\tilde{\omega}, R_d, \omega_d) \in \mathcal{W}_z$, respectively. In the sequel, the tracking objective is reformulated as to stabilize the compact set $\{z \in \mathcal{W}_z: \tilde{R} = I, \tilde{\omega} = 0\}$ for the plant, of which the dynamics is described by the set-valued map $f: \mathcal{W}_z \times \mathbb{R}^3 \rightrightarrows \mathcal{M}_z \coloneqq \mathbb{R}^{3\times 3} \times \mathbb{R}^3 \times \mathbb{R}^{3\times 3} \times \mathbb{R}^3$ defined as
\begin{equation} \label{eq:plant}
    \dot{z} \in f(z, \tau) \coloneqq
  \begin{pmatrix}
    \tilde{R} (R_d \tilde{\omega})^\wedge\\
    \Sigma ( \tilde{\omega} , \omega_d) \tilde{\omega} - \Phi(\tilde{\omega},\omega_d,\dot{\omega}_d) + \tau \\
    R_d  \omega_d^\wedge \\
    \bar{\mathbb{B}}^3_{c_a}
  \end{pmatrix} .
\end{equation}
Now, the following hybrid feedback control scheme is proposed
\begin{equation} \label{eq:controller_1}
  \mathcal{H}_{K} : \ 
  \begin{cases}
    \dot{q} = 0 & (q,z) \in \mathcal{F}_{1} ,\\
    q^+ \in G_{K} (q,z) & (q,z) \in \mathcal{J}_{1}, \\
    \tau = \kappa_1 (q,z).
  \end{cases}
\end{equation}
The flow and jump sets are defined as $\mathcal{F}_{1}  = \{ (q,z) \in \mathcal{Q} \times \mathcal{W}_z : \pi_V (\tilde{R},q) \leq \delta(q) \}$ and $\mathcal{J}_{1}  = \{ (q,z) \in  \mathcal{Q} \times \mathcal{W}_z : \pi_V (\tilde{R},q) \geq \delta(q) \}$. The jump map $G_{K} : \mathcal{Q} \times \mathcal{W}_z \rightrightarrows \mathcal{Q} $ is given by $G_{K}  (q,z) =   \operatorname{argmin}_{p \in \mathcal{Q}} V(\tilde{R},p) $. The geometric nonlinear PD control law $\kappa_1 : \mathcal{Q} \times \mathcal{W}_z \to \mathbb{R}^3 $ is defined as 
\begin{equation}\label{eq:ctr_law1}
    \kappa_1 (q,z) = \Phi(\tilde{\omega} ,\omega_d,\dot{\omega}_d) - k_1 R_d^\top  \rho_V(\tilde{R},q) - k_2\tilde{\omega}
\end{equation}
where $k_1, k_2 > 0$ are positive gains and the three terms represent the feedforward action, proportional action, and derivative action, respectively. We refer to \cite{Bullo1999,Bullo2004,Maithripala2015} for further study of geometric PD controllers for mechanical systems.

\begin{figure}[t]
  \centering
  \includegraphics[width = 8cm]{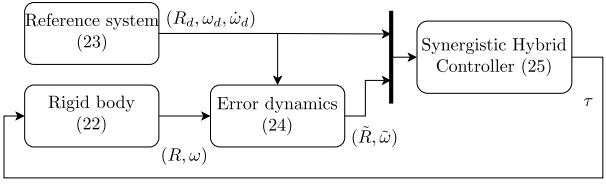}
  \caption{The block diagram of the closed-loop system for attitude tracking.}
  \label{fig:blk_diag_AttTrack}
\end{figure}

It has been established in \cite{Mayhew2013,Casau2015,Berkane2017b,Wang2021} that the control law \eqref{eq:ctr_law1} renders global asymptotic tracking and global exponential tracking in the common used synergistic control where the synergy condition is described by using $\mu_V$; see Figure~\ref{fig:flowchart}. The primary difference between our approach and those presented in literature lies in our refined synergistic control given by Lemma \ref{lem:cenSynergyLocal}, yet the global stability of the closed-loop therein still hold. This can save computation cost because only selected potential functions $\{V(X,q): q \in \mathcal{Q}_q\}$ instead of $\{V(X,q): q \in \mathcal{Q}\}$ need to be evaluated at each update.  

Let us define the new state space $\mathcal{X}_1 = \mathcal{Q} \times \mathcal{W}_z$ and new state $x_1 = (q,z) \in \mathcal{X}_1$. In the sequel, the hybrid closed-loop system resulted from the plant \eqref{eq:plant} and hybrid controller \eqref{eq:controller_1} is represented as
\begin{equation} \label{eq:hyb_clp1}
    \mathcal{H}_1 : \
    \begin{cases}
        \dot{x}_1 \in F_1 (x_1) & x_1 \in \mathcal{F}_{1} ,\\
        x_1^+ \in G_1 (x_1) & x_1 \in \mathcal{J}_{1} ,
    \end{cases}
\end{equation}
where the flow map $F_1 : \mathcal{X}_1  \rightrightarrows \mathbb{R} \times \mathcal{M}_z  $ and jump map $G_1 : \mathcal{X}_1 \rightrightarrows \mathcal{X}_1 $ are defined as $F_1 (x_1) = (0, f(z,\kappa_1(q,z)))$ and $G_1 (x_1) = (G_{K}  (q,z) , z)$, where $f$ is given by \eqref{eq:plant} and $G_K$ is defined in \eqref{eq:controller_1}. The block diagram of the hybrid closed-loop system \eqref{eq:hyb_clp1} is plotted in Figure~\ref{fig:blk_diag_AttTrack}. Of note, the system \eqref{eq:hyb_clp1} is autonomous and satisfies the hybrid basic conditions \cite{Goebel2009}. The next proposition states that the global tracking is achieved under the system \eqref{eq:hyb_clp1}.

\begin{prop} \label{prop:stability}
  Let $k_1, k_2 > 0$. Then the compact set $\mathcal{A}_1 \coloneqq \{x_1 \in \mathcal{X}_1 : \tilde{R} = I, \tilde{\omega} = 0\}$ is robustly and globally asymptotically stable for the closed-loop system \eqref{eq:hyb_clp1} and the number of jumps is finite. 
\end{prop}

\section{Simulations} \label{sec:sim}

In this section, we present a simulation study to validate the central synergy of the synergistic potential functions \eqref{eq:warpMPF} and its application to attitude tracking. 

\subsection{Example: synergistic function on $\mathrm{SO(3)}$}
Suppose that there are three inertial vectors, $a_i = e_i,\; i = 1,2,3$ where $e_i \in \mathbb{R}^3$ is the $i$-th vector of $I_3$. Let the weights be $w_1 = 0.2, w_2 = w_3 = 0.4$. By \eqref{eq:wahba}, it follows that $M = \mathrm{diag}([0.2,0.4,0.4])$ and subsequently, $G = \mathrm{diag}([0.6,0.6,0.8])$ which is used in the standard modified trace function. Let us observe the behavior of the synergistic family at the typical unwanted critical points at which $V(X,q) = 2\lambda_2^G$. 

\begin{figure}[ht]
    \centering
    \includegraphics[width = 8cm]{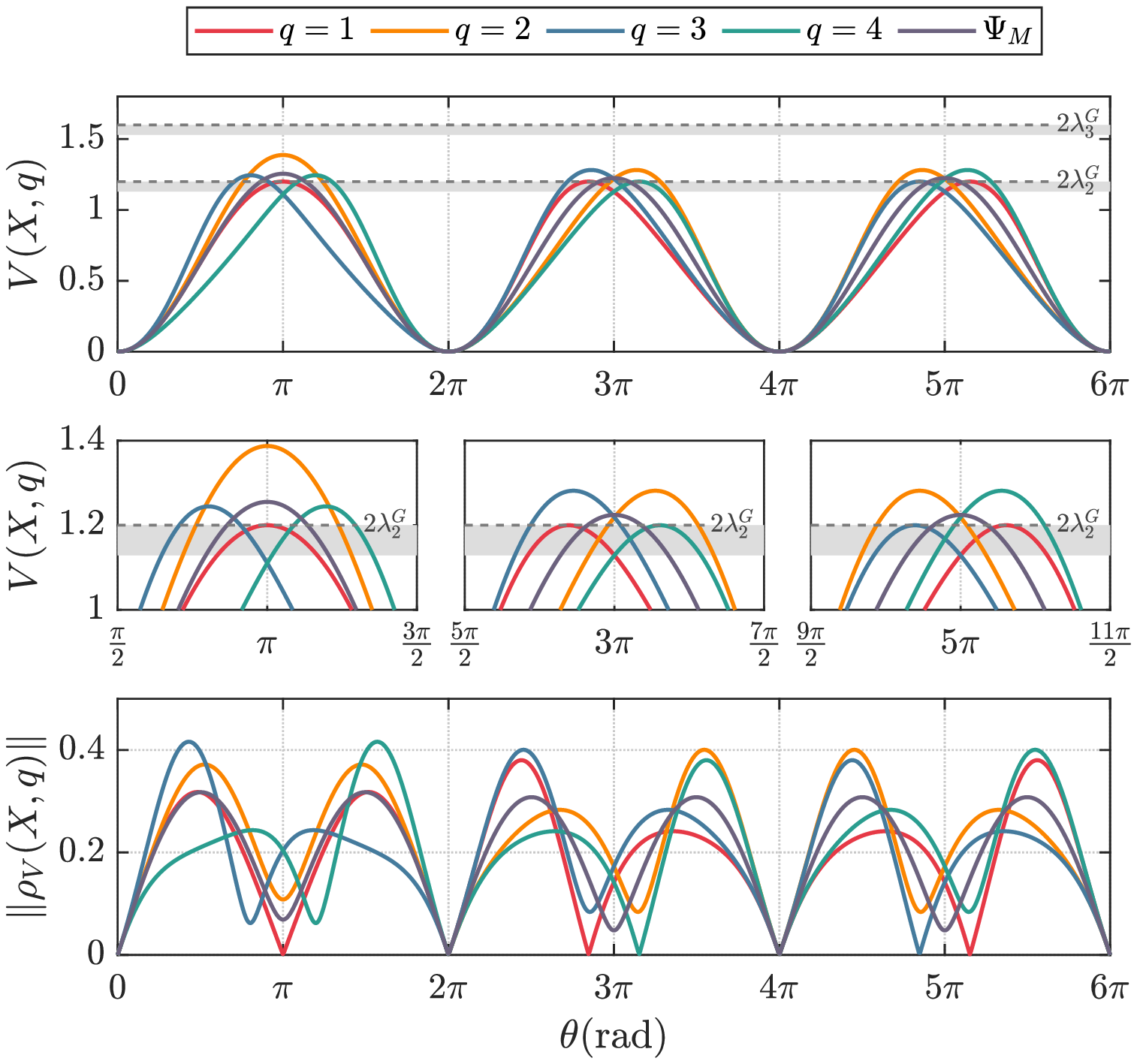}
    \caption{Synergistic potential functions of the item (2) in Theorem \ref{thm:SelectDirctions}. The input attitude $X = \mathcal{R}_a(\theta, u)$ where $u = [0.37, 0, 0.93]^\top $ for $\theta \in [0,2\pi]$, $u = [0.25, 0.69, 0.69]^\top $ for $\theta \in [2\pi,4\pi]$, and $u = [0.25, -0.69, 0.69]^\top $ for $\theta \in [4\pi,6\pi]$. The dash y-line is $V(X,q) = 2 \lambda^G$ and the shadow area is $2 \lambda^G \leq V(X,q) \leq 2 \lambda^G - \bar{\delta}_q$.} 
    \label{fig:synMpfType2}
\end{figure}

First, we take consider the item (2) in Theorem \ref{thm:SelectDirctions}. It follows that $\mathcal{Q} = \{1,2,3,4\}$ and the warping directions are given by $\mathcal{U} = \{e_2, -e_2, e_3, -e_3\}$. Let $k $ defined in \eqref{eq:WarpAngle_MPF} be $ 0.465$ and subsequently $\bar{\delta}_q = 0.0712$ by Theorem \ref{thm:explicit_syngap}. Figure~\ref{fig:synMpfType2} illustrates the evaluation of $V(X,q)$ and $\Psi_M(X)$ at $X = \mathcal{R}_a(\theta , u)$ for some specific $u$. Each $V(X,p)$ for $p \in \{2,3,4\}$ may exceed $V(X,1)$ at the critical points of $V(X,1)$ belonging to $\mathcal{T}^{-1}(\mathcal{R}_a(\pi , e_2))$. Moreover, it verifies the results from Theorem \ref{thm:SelectDirctions} and \ref{thm:explicit_syngap} that, $V(X,p) < V(X,1)$ always hold for either $p = 3$ or $p = 4$ at $\operatorname{Crit} (V(X,1)) \setminus \mathcal{A}$ and the refined synergistic gap $\pi_V(X,q)$ definitely exceeds the ${\delta}_q$ at its unwanted critical points. 

\begin{figure}[ht]
  \centering
  \includegraphics[width = 8cm]{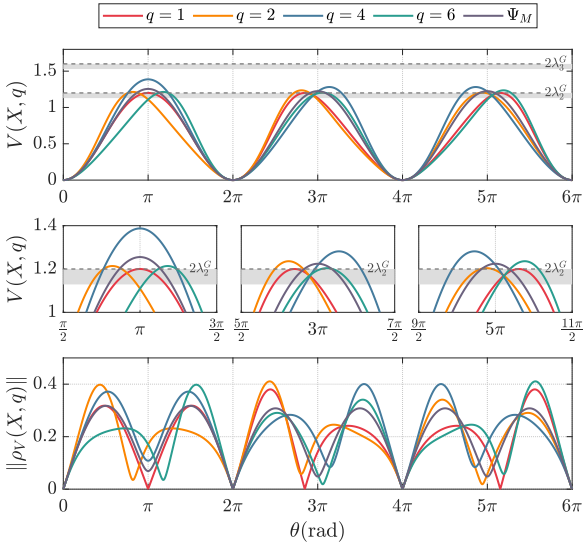}
  \caption{Synergistic potential functions of the item (3) in Theorem \ref{thm:SelectDirctions}. The input attitude $X = \mathcal{R}_a(\theta, u)$ where $u = [0.37, 0, 0.93]^\top $ for $\theta \in [0,2\pi]$, $u = [0.25, 0.69, 0.69]^\top $ for $\theta \in [2\pi,4\pi]$, and $u = [0.25, -0.69, 0.69]^\top $ for $\theta \in [4\pi,6\pi]$. The dash y-line is $V(X,q) = 2 \lambda^G$ and the shadow area is $2 \lambda^G \leq V(X,q) \leq 2 \lambda^G - \bar{\delta}_q$.} 
  \label{fig:synMpfType3}
\end{figure}

Next, we consider the item (3) in Theorem \ref{thm:SelectDirctions} and, in consequence, $\mathcal{Q} = \{1,\dots,6\}$ and the warping directions $\mathcal{U} = \{e_2 \cos (\frac{n\pi}{3})+e_3 \sin(\frac{n \pi}{3}): n = 0,\dots,5\}$. Continuing with $k = 0.465$, we set $\bar{\delta}_q = 0.0712$ by Theorem \ref{thm:explicit_syngap}. Letting $q = 1$ and $u_1 = e_2$, we observe the functions for $p \in \mathcal{Q}_1 = \{2,4,6\}$. As shown in Figure~\ref{fig:synMpfType3}, there always exists $p \in \mathcal{Q}_1$ such that $V(X,1) - V(X,p) \geq \bar{\delta}_q$ for $(X,1) \in \operatorname{Crit} (V(X,1)) \setminus \mathcal{A}$, which is consistent with Theorems \ref{thm:SelectDirctions} and \ref{thm:explicit_syngap}.

\subsection{Examples: attitude tracking}
The proposed centrally synergistic hybrid controller \eqref{eq:controller_1} is referred to as `$\pi_V$-CS'. Three controllers are implemented for comparison. The first controller is from \eqref{eq:ctr_law1} with $q \equiv 1$ (i.e., without the switching mechanism) and is referred to as `Solo'. The second controller is \eqref{eq:controller_1} with the traditional switching mechanism (i.e., the flow and jump sets in \eqref{eq:controller_1} are defined with $\mu_V$ instead of $\pi_V$) and is referred to as `$\mu_V$-CS'. The third controller is the non-centrally synergistic hybrid controller of \eqref{eq:ctr_nonCS} in \cite{Lee2015} and is referred to as `NonCS'.

Consider a rigid body, with an inertia matrix of $J = \mathrm{diag}([0.5,0.7,0.3])\si{\kilogram.\metre^2}$, being required to track a trajectory given by $R_d(0) = I$ and $\omega_d(t) = [te^{-0.5t}, 0.6\sin(0.4t), 0.6\sin(0.7t)]^\top $. The sampling interval is set as $1\si{ms}$. The measurement of rigid-body attitude is given by $\bar{R} = R \mathcal{R}_a(\alpha_r, n_r) $, where $n_r = n/|n|$, each element of $n$ is normally distributed with unit standard deviation, and $\alpha_r$ was drawn from a uniform distribution on the interval $(0, 0.01\pi)$. The measurement of the angular rate is given by $\bar{\omega} = \omega + n_\omega$, where $n_\omega$ is normally distributed with standard deviation $0.01$. Suppose that $M = \mathrm{diag}([0.2,0.4,0.4])$ and the synergistic functions used in Solo, $\pi_V$-CS, and $\mu_V$-CS were constructed by using the case (2) of Theorem \ref{thm:SelectDirctions} and \ref{thm:explicit_syngap} in the preceding section, where $\delta(q)$ is set as $0.8\bar{\delta}_q = 0.057$. Note that the centrally synergistic functions cannot be generated from the modified trace function in this case by using the approaches in \cite{Mayhew2011b,Berkane2017b,Berkane2017a,Casau2015,Wang2021}, because they require either $M$ with distinct eigenvalues or $M$ with repeated maximum eigenvalues. The parameters of the controllers are determined empirically as follows. The gains of Solo, $\pi_V$-CS, and $\mu_V$-CS are $k_1 = 60$, $k_2 = 6$. For NonCS, $k_1 = 30$, $k_2 = 3$, $\alpha = 1.5$, $\beta = 0.4$, and $\delta = 0.025$.

Each figure has the six plots: the attitude tracking error $\vartheta (\tilde{R}) = \arccos \bigl(\frac{1}{2}(\operatorname{tr} ( \tilde{R} )-1 ) \bigr)$; the velocity tracking error $|\tilde{\omega}|$; the norm of the input torque $|\tau|$; the logical variable $q$; the synergistic function $V(\tilde{R},q)$; $\int N (t) dt$ representing the computational complexity of checking the switching condition, where $N$ denotes the counts of evaluating the synergistic functions at each update.

\begin{figure}[t]
  \centering
  \includegraphics[width = 8cm]{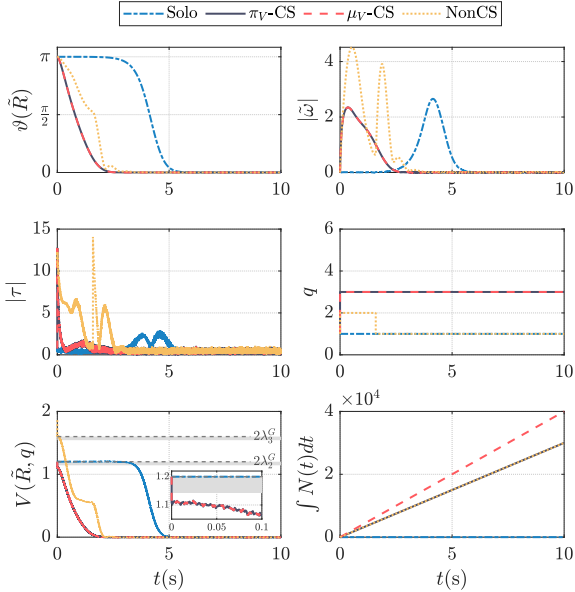}
  \caption{Initial conditions: $R(0) = \mathcal{R}_a(\pi , [0.37, 0, 0.93]^\top )$, $\omega(0) = 0$ and $q = 1$. The Solo stopped at the unwanted critical points, but the hybrid controllers did not.}
  \label{fig:trackType2_SlowCR}
\end{figure}

As shown in Figure~\ref{fig:trackType2_SlowCR}, the state feedback of the Solo vanishes at the onset since $\rho_V(\tilde{R},q) = 0$ at its critical points. Consequently, the attitude error $\tilde{\theta}$ remains at $\pi$ for some time, although it finally converges to zero owing to the instability of the initial critical point. By comparison, the CS hybrid controllers are invulnerable to the unwanted critical points, because Theorem \ref{thm:explicit_syngap} guarantees the existence of a smaller potential function when the current one is trapped in its unwanted critical points. Particularly, the proposed $\pi_V$-CS method can save the computation cost in the sense that the number of the potential functions to be evaluated at each update is decreased. In addition, the Non-CS hybrid controller can also avoid the unwanted critical points. In a word, the synergistic hybrid controllers exhibited an improved convergence rate of the tracking error than the continuous controller. 

\begin{figure}[t]
  \centering
  \includegraphics[width = 8cm]{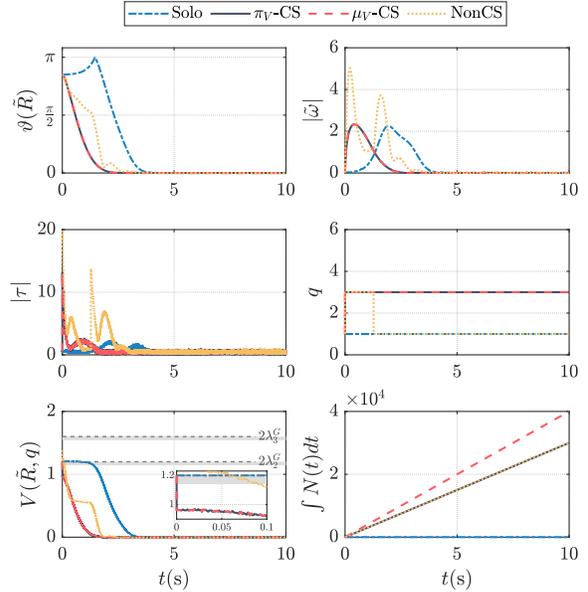}
  \caption{Initial conditions: $R(0) = \mathcal{R}_a(1.15\pi, [0.25,-0.69,0.69]^\top )$, $\omega (0) = 0$, and $q = 1$. The Solo exhibited the unwinding phenomenon, but the hybrid controllers avoided it.}
  \label{fig:trackType2_Unwinding}
\end{figure}

Figure~\ref{fig:trackType2_Unwinding} illustrates that the Solo exhibited the unwinding phenomenon, i.e., that the feedback pulls the potential function toward zero but in a longer direction. The hybrid controllers can avoid this undesired phenomenon.

\begin{figure}[t]
  \centering
  \includegraphics[width = 8cm]{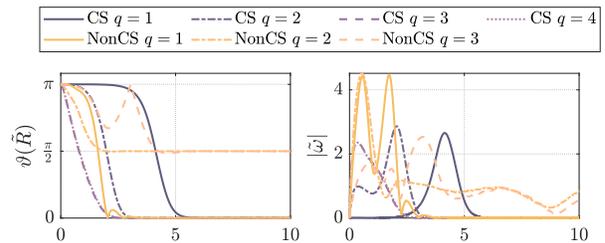}
  \caption{Hybrid controllers without switching mechanisms and the initial conditions: $R(0) = \mathcal{R}_a(\pi , [0.37, 0, 0.93]^\top )$ and $\omega(0) = 0$. }
  \label{fig:trackType2_CSvsNonCS}
\end{figure}

Finally, in order to observe the difference between our CS hybrid controller and the Non-CS hybrid controller, we compare the control laws in $\pi_V$-CS and NonCS without the switching mechanism. As shown in Figure~\ref{fig:trackType2_CSvsNonCS}, each control law in $\pi_V$-CS can obtain the tracking task individually, although they can suffer from the slow convergence rate due to the existence of the undesired critical points. However, not all the control laws in NonCS can accomplish attitude tracking, in particular the NonCS for $q=2$ and $q=3$ stabilized the attitude tracking error at a constant rotation angle. Therefore, the centrally synergistic design is more robust to the malfunction of switching mechanism than the non-synergistic design.

\section{Conclusion} \label{sec:con}
In this paper, we demonstrated that centrally synergistic potential functions on $\mathrm{SO(3)}$ can be generated from a single modified trace functions by angular warping in multiple directions. Interestingly, although the amount of synergistic functions increases, synergistic control is effective in such a way that the switching mechanism is operating within a specific subset for each function of the family, which is coined ``refined synergistic control" herein. In addition, the positive lower bound of synergistic gap was determined explicitly by use of the warping angle function in the literature. The proposed synergistic functions were applied to attain robust global attitude tracking and illustrated by numerical examples. Possible future directions of this research include but are not restricted to the design of the centrally synergistic control using the unit-quaternion-based feedback and the extension of the proposed centrally synergistic control to the state space of the rigid-body pose.

\appendix \label{sec:app}

\section{Proof of Lemma \ref{lem:cenSynergyLocal}}  \label{pf:lem_cenSynergyLocal}
\begin{pf}
  According to Definition \ref{def:pfunc}, the potential function is continuously differentiable and attains its global minimum at $I$. Hence, $I$ is a critical point for all potential functions. That is to say, $\mathcal{A} \subset \operatorname{Crit} (V) $. This shows the item (1). 

  Necessity of the item (2) is straightforward because \eqref{eq:cenSynergy} implies \eqref{eq:cenSynergyLocal} if we set $\mathcal{Q}_q \equiv \mathcal{Q}$ and $\delta(q) \equiv \delta$. Now, let us illustrate the sufficiency. Since $\min_{p\in \mathcal{Q}} V(X,p) \leq \min_{p\in \mathcal{Q}_q} V(X,p)$ for all nonempty subset $\mathcal{Q}_q \subset \mathcal{Q}$, it follows immediately that $\mu_V(X,q) \geq \pi_V (X,q) > \delta(q) > 0$ for all $(X,q) \in  \operatorname{Crit} (V) \setminus \mathcal{A}$. Therefore, setting $\delta \coloneqq \min_{q\in \mathcal{Q}} \delta(q)$ yields \eqref{eq:cenSynergy}.  \qed
\end{pf}

\section{Proof of Lemma \ref{lem:condSynergMPF}} \label{pf:lem_condSynergMPF}
\begin{pf}
  According to \eqref{eq:critMPF} and Lemma \ref{lem:synergyCon}, the set of critical points of $V$ is given by $\operatorname{Crit} (V) = \mathcal{A} \cup_{v\in \mathcal{E}(M)} \mathcal{T}^{-1} (\mathcal{R}_a(\pi,v))$. This shows the item (1).
  
  Since $(Y,q) \in \operatorname{Crit} (V) \setminus \mathcal{A}$ and $\theta$ is positive definite with respect to $\{I\}$, we have that $Y \neq I$ and hence that $0 < \theta(Y) < \pi$. Making use of the quaternion multiplication, the quaternion scalar part of $\mathcal{R}_a(\theta_{pq}, u_{pq})$ is given by \eqref{eq:theta_pq_def}. In view of ${u_p^\top }{u_q} \in [-1,1) $ and $\theta(Y) \in (0,\pi)$, one has that $ -1 < \cos(\frac{\theta_{pq}}{2}) < 1$. Thereupon, it follows that $\sin^2 (\frac{\theta_{pq}}{2}) > 0$ and $u_{pq}$ given by \eqref{eq:u_pq_def} is well-defined. This shows the item (2).
  
  We now turn to the item (3). For each $ v\in \mathcal{E}(M) $, we can write $(Y,q)$ and $V(Y,q)$ as 
  \begin{equation*}
    \begin{split}
        Y &= \mathcal{R}_a(\pi, v) \mathcal{R}_a(\theta(Y), u_q)^\top ,\\
        V(Y,q) &= \Psi_M (\mathcal{R}_a(\pi, v)) = 2 v^\top  G v = 2 \lambda^G,
    \end{split}
  \end{equation*}
  where \eqref{eq:EvaModPF} is used.
  For a given $p \in \mathcal{Q} \setminus \{q\}$, substituting $(Y ,p)$ back to \eqref{eq:angularTrans} yields
  \begin{align*}
      \mathcal{T}(Y,p) &= Y \mathcal{R}_a(\theta (Y) , u_p)  = \mathcal{R}_a(\pi , v) \mathcal{R}_a(\theta_{pq} , u_{pq}) 
  \end{align*}
  where $u_{pq}$ and $\theta_{pq}$ is given by \eqref{eq:u_pq_def} and \eqref{eq:theta_pq_def}, respectively. By \eqref{eq:EvaModPFPi} and Lemma \ref{lem:cenSynergyLocal}, if $V$ is centrally synergistic relative to $\mathcal{A}$, then there exists a subset $\mathcal{Q}_q \subseteq \mathcal{Q}$ for each $q \in \mathcal{Q}$ such that $\pi_V (Y,q) = \max_{p \in \mathcal{Q}_q \setminus \{q\}}  (1- \cos ({\theta_{pq}})) \Delta(v,u_{pq}) > 0$. This implies \eqref{eq:cond_synerg} because $1- \cos ({\theta_{pq}}) =2 \sin^2 (\frac{\theta_{pq}}{2}) > 0$ from the item (2). This shows the item (3) and completes the proof. \qed
\end{pf}

\section{Proof of Theorem \ref{thm:SelectDirctions}} \label{pf:thm_SelectDirctions}
\begin{pf}
  There is no loss of generality in assuming $v_3 = v_1 \times v_2$. By Lemma \ref{lem:condSynergMPF}, for all $v \in \mathcal{E}(M)$, $(Y,q) \in \mathcal{T}^{-1} (\mathcal{R}_a(\pi , v))$ is the related unwanted critical point of $V$ and satisfies $0 < \theta(Y) < \pi$. To prove \eqref{eq:cond_synerg}, we need to calculate $\Delta(v,u_{pq})$. Note that the $u_{pq}$ given by \eqref{eq:u_pq_def} and \eqref{eq:theta_pq_def} is associated to the sign of $\sin (\theta_{pq}/2) $. For simplicity, we only consider the case of $\sin (\theta_{pq}/2) > 0$, since $\Delta(v,u_{pq}) = \Delta(v,-u_{pq})$ for all $v,u_{pq} \in \mathbb{S}^2$. 

  \setcounter{case}{0}
  \begin{case}[$\lambda_i^M = \lambda >0, \ i  = 1,2,3 $]
    In this case, the set of the unit eigenvectors of $M$ is given by $\mathcal{E}(M) = \mathbb{S}^2$. By Lemma \ref{lem:property_MPF}, one can write \eqref{eq:cond_synerg} as 
    \begin{equation}
      \max_{p \in \mathcal{Q}_q \setminus \{q\}} \lambda^G (v^\top  u_{pq})^2 > 0 . \label{eq:equiv_condSynergy1}
    \end{equation}
    If $u_q = v_1$, which we may assume, then $\mathcal{U}_q = \{\pm v_2, \pm v_3 \}$. Computing \eqref{eq:u_pq_def} and \eqref{eq:theta_pq_def} for all  $u_p \in \mathcal{U}_q$ to  yields 
    \begin{align} 
      & u_{pq} = 
      \begin{cases}
        \frac{ (- v_1 \pm v_2  ) \cos (\frac{\theta(Y)}{2}) \mp v_3  \sin (\frac{\theta(Y)}{2})}{\sqrt{1 + \cos^2 (\frac{\theta(Y)}{2})}}, &  (u_p = \pm v_2),\\
        \frac{ ( - v_1 \pm v_3  ) \cos (\frac{\theta(Y)}{2}) \pm v_2  \sin (\frac{\theta(Y)}{2})}{\sqrt{1 + \cos^2 (\frac{\theta(Y)}{2})}}, &  (u_p = \pm v_3),
      \end{cases} \label{eq:u_pq_perp} \\
      & \sin ( {\theta_{pq}}/{2}) = \sin \bigl( {\theta(Y)}/{2} \bigr) \sqrt{1 + \cos^2 \bigl( {\theta(Y)}/{2}\bigr)}. \label{eq:theta_pq_perp}
    \end{align}
    Through computing the Jacobian determinant of the first three elements of $\{u_{pq} \}$, one can show that the set $\{u_{pq} \in \mathbb{R}^3: u_p \in \mathcal{U}_q \}$ spans $\mathbb{R}^3$ for $\theta(Y ) \in (0,\pi)$. Consequently, \eqref{eq:equiv_condSynergy1} holds for all $v \in \mathcal{E}(M)$ with the postulated condition $u_q = v_1$. By contraposition, the conclusion is not affected for all $u_q \in \mathcal{U} $. This shows the item (1).
  \end{case}
  \begin{case}[$\lambda_1^M = \lambda^M_2 > \lambda^M_3 \geq 0$]
    In this case, $\mathcal{E}(M) =\{  v_3, - v_3  \} \cup \{v_{12} = v_1 \cos (t) + v_2 \sin(t): t \in \mathbb{R} \}  $ and hence \eqref{eq:cond_synerg} is divided into two possible cases: $v = v_3$ and $v = v_{12}$, respectively, 
    \begin{subequations}
      \begin{align}
        \max_{p \in \mathcal{Q}_q \setminus \{q\}} \lambda_3^G - \lambda_2^G (1 - (u^\top _{pq} v_3)^2) & > 0 , \label{eq:equiv_condSynergy2a} \\
        \max_{p \in \mathcal{Q}_q \setminus \{q\}}  (1 - (u^\top _{pq} v_3)^2) (\lambda_2^G - \lambda_3^G \sin^2 (\phi_p)) & >0, \label{eq:equiv_condSynergy2b} 
      \end{align}
    \end{subequations}
    where $\phi_p = \angle(v_{12},u_{pq}^\perp)$ and $u_{pq}^\perp = (I - v_3 v_3^\top ) u_{pq}$. In view of \eqref{eq:eigGM}, we have that $\lambda_3^G - \lambda_2^G = \lambda_2^M - \lambda_3^M > 0$ and in consequence, that \eqref{eq:equiv_condSynergy2a} holds for all $p \in \mathcal{Q}\setminus \{q\}$. It is reduced to show \eqref{eq:equiv_condSynergy2b} with $(Y,q) \in \mathcal{T}^{-1} \mathcal{R}_a(\pi , v_{12})$.

    Consider the item (2) and suppose that $u_q = v_1$. It follows that $\mathcal{U}_q = \{\pm v_2\}$ and that $u_{pq}$ and $\theta_{pq}$ are given by \eqref{eq:u_pq_perp} and \eqref{eq:theta_pq_perp}, respectively. It is easy to check that 
    \begin{equation}
      (u_{pq}^\top  v_3)^2 = \sin^2 ( {\theta_{pq}}/{2}),\ \forall \theta(Y) \in (0,\pi). \label{eq:u_pq_dot_v3_perp}
    \end{equation}
    Moreover, we have that
    \begin{align}  \label{eq:u_pq_perp_perp}
       {u_{pq}^{\perp}}/{|u_{pq}^{\perp}|} &=   ( - v_1 \pm v_2) / {\sqrt{2}}, & (u_p = \pm v_2) . 
    \end{align}
    Setting $v_{12} = v_1 \cos (t ) + v_2 \sin (t)$ with $t \in \mathbb{R}$ yields  
    \begin{equation*}
      \min_{u_p \in \mathcal{U}_q}\sin^2( \phi_p ) = ( 1 - |\sin(2t)| ) /2 <  {\lambda_2^G}/{\lambda_3^G}, \label{eq:phi_p_perp}
    \end{equation*}
    where the inequality follows from the fact that $\lambda_2^G / \lambda_3^G > \lambda_1^M  / (\lambda_1^M + \lambda_2^M) = \frac{1}{2}$. This formula and \eqref{eq:u_pq_dot_v3_perp} implies that \eqref{eq:equiv_condSynergy2b} holds for all $v_{12} \in \mathrm{span}\{v_1,v_2\} \cap \mathbb{S}^2$ with the postulated condition $u_q = v_1$. Additionally, by contraposition, the conclusion is not affected for all $u_q \in \mathcal{U} $. This shows the item (2).

    Consider the item (3) and assume that $u_q = v_1$. It follows that $\mathcal{U}_q = \{- v_1, \frac{1}{2} v_1 + \frac{\sqrt{3}}{2} v_2,\frac{1}{2} v_1 - \frac{\sqrt{3}}{2} v_2\}$. Applying $\mathcal{U}_q$ to \eqref{eq:u_pq_def} and \eqref{eq:theta_pq_def} yields
    \begin{equation} \label{eq:u_pq_pi3}
      \begin{split}
        & u_{pq} = \\
        & \begin{cases}
          -v_1, & (u_p = -v_1), \\
          \frac{(- \frac{1}{2} v_1 \pm \frac{\sqrt{3}}{2} v_2  ) \cos (\frac{\theta(Y)}{2}) \mp \frac{\sqrt{3}}{2} v_3 \sin (\frac{\theta(Y)}{2})}{ \frac{1}{2} \sqrt{3 + \cos^2 (\frac{\theta(Y)}{2})}  }, & (u_p = \frac{v_1 \pm \sqrt{3}v_2}{2}) , 
        \end{cases}
      \end{split}
    \end{equation}
    \begin{equation} \label{eq:theta_pq_pi3}
      \sin (\frac{\theta_{pq}}{2}) =  
        \begin{cases}
          \sin (\theta(Y)), & (u_p = -v_1), \\
          \frac{\sqrt{3 + \cos^2 (\frac{\theta(Y)}{2})}}{2} \sin (\frac{\theta(Y)}{2})  , & (u_p = \frac{v_1 \pm \sqrt{3}v_2}{2}) .
        \end{cases}
    \end{equation}
    A straightforward calculation yields
    \begin{equation} \label{eq:u_pq_dot_v3_3pi}
      (u_{pq}^\top v_3)^2 =  
        \begin{cases}
          0, & (u_p = -v_1), \\
          \frac{3\sin^2(\frac{\theta(Y)}{2})}{3 + \cos^2(\frac{\theta(Y)}{2}) }, & (u_p = \frac{v_1 \pm \sqrt{3}v_2}{2}) .
        \end{cases}
    \end{equation}
    It is easily to see $(u_{pq}^\top v_3)^2 < 1$ for all $\theta(Y) \in (0,\pi)$. Moreover, we have
    \begin{equation} \label{eq:u_pq_perp_3pi}
      \frac{u_{pq}^{\perp}}{|u_{pq}^{\perp}|} =
      \begin{cases}
        -v_1, & (u_p = -v_1), \\
        - \frac{1}{2} v_1 \pm \frac{\sqrt{3}}{2} v_2  , & (u_p = \frac{v_1 \pm \sqrt{3}v_2}{2}) . 
      \end{cases}
    \end{equation}
    Employing $v_{12} = v_1 \cos (t ) + v_2 \sin (t)$ with $t \in \mathbb{R}$ yields 
    \begin{align*}
      \min_{u_p \in \mathcal{U}_q} \sin^2 (\phi_p) &= \min \left\{\sin^2 (t),\sin^2 (t + \frac{\pi}{3}),\sin^2 (t - \frac{\pi}{3})\right\} . 
    \end{align*}
    Since $\min_{u_p \in \mathcal{U}_q} \sin^2 (\phi_p)\leq \frac{1}{4} < \lambda_1^M  / (\lambda_1^M + \lambda_2^M) \leq \lambda_2^G/\lambda_3^G$ for all $t\in \mathbb{R}$, it follows that \eqref{eq:equiv_condSynergy2b} holds for all $v_{12} \in \mathrm{span}\{v_1,v_2\} \cap \mathbb{S}^2$ with the postulated condition $u_q = v_1$. The conclusion involves no loss of generality for all $u_q \in \mathcal{U} $. This shows the item (3).
    
  \end{case}

  \begin{case}[$0 < \lambda_1^M = \lambda_2^M < \lambda_3^M$ or $0 \leq \lambda_1^M < \lambda_2^M < \lambda_3^M$]

    The items (4-5) can be verified by \eqref{eq:EvaModPFPi} and Lemma \ref{lem:condSynergMPF}.
  \end{case} 
  This finishes the proof. \qed
\end{pf}

\section{Proof of Theorem \ref{thm:explicit_syngap}} \label{pf:thm_explicit_syngap}
The following result will be used to prove Theorem \ref{thm:explicit_syngap}.
\begin{prop} \label{prop:min_F}
  Consider the following real-valued functions on $\mathbb{R}$, $f_1 (t) = 4(\xi - \sin^2 (t) ) $, $ f_2 (t) = \xi - \sin^2 (t + \frac{\pi}{3})$, $ f_3 (t) = \xi - \sin^2 (t - \frac{\pi}{3})$, and $F(t ) = \max \{f_1(t), f_2(t),f_3(t)\}$, where $\xi \in [\frac{1}{2},1]$ is constant. Then, $\min_{t \in \mathbb{R} } F(t) = \xi - \frac{1}{4}$. 
\end{prop}
\begin{pf}
  Since $\min_{t \in \mathbb{R}}  \{\sin^2 (t),\sin^2 (t + \frac{\pi}{3}),\sin^2 (t - \frac{\pi}{3}) \}  \leq \frac{1}{4}$, one can assert that $F (t) > 0$ for all $t \in \mathbb{R} $. Making use of the fact that $F(t) = F(t + \pi)$ and $F(t) = F(\pi - t) $ for each $t \in \mathbb{R}$, one obtains that $\min_{t \in \mathbb{R}} F(t) = \min_{t \in [0, \frac{\pi}{2}]} \bigl( \max \{f_1(t), f_3(t)\} \bigr)$. Define $g(t) = f_1(t) - f_3(t)$, then it follows from basic rules of calculus that $g$ is strictly decreasing on $[0, \frac{\pi}{3}]$. Since $g(0) = 3 \xi + \frac{3}{4} > 0$ and $g(\frac{\pi}{3}) = 3 \xi - 3  \leq  0$, one can assert that $g$ has only one x-intercept on $[0, \frac{\pi}{3}]$. Using the Matlab Symbolic Computation Tool, the solution to $g (t_0) = 0$ subject to $ 0. 5 \leq \xi \leq 1$ and $ t_0 \in [0, \frac{\pi}{3}]$ is given by $\sin (t_0) = \sqrt{ \frac{9\xi}{14} - \frac{\sqrt{-3 \xi^2 + 3\xi + 1}}{14} + \frac{5}{28}}$. Therefore, $F(t) = f_1 (t)\geq f_1(t_0) $ on $[0,t_0]$ since $g(t) \geq 0$. On the other hand, one can verify that $t_0 $ is strictly increasing with $\xi \in [\frac{1}{2},1]$, so it gives $\sin(t_0) \geq \sqrt{\frac{14 - \sqrt{7}}{28}} > \frac{1}{2}$ and $t_0  >  \frac{\pi}{6}$. Seeing that $f_3$ is concave on $[\frac{\pi}{6},\frac{\pi}{2}]$, one has $\min_{t \in [t_0,\frac{\pi}{2}]} f_3(t) = f_3(\frac{\pi}{2})$, where $f_3(\frac{\pi}{2}) = f_3(\frac{\pi}{6}) < f_3(t_0)$ is used. Since $ F(t) \geq f_3(t) \geq f_3(\frac{\pi}{2})$ for each $t \in [t_0, \frac{\pi}{2}]$ and $F(\frac{\pi}{2}) = f_3(\frac{\pi}{2})$, one can assert $\min_{t \in [t_0,\frac{\pi}{2}]} F(t) = f_3(\frac{\pi}{2})$. In view of $f_3(\frac{\pi}{2}) < f_3(t_0) = f_1(t_0)$, it follows that $\min_{t \in \mathbb{R} } F(t) = \min \bigl\{f_1(t_0), f_3(\frac{\pi}{2})\bigr\} = \xi - \frac{1}{4} $, which completes the proof. \qed 
\end{pf}

\begin{pf}
  We first compute the unwanted critical point $(Y,q) \in \operatorname{Crit} (V) \setminus \mathcal{A}$. By Lemma \ref{lem:condSynergMPF}, $Y  = \mathcal{R}_a(\pi , v) \mathcal{R}_a(\theta(Y) , u_q)^\top$ for some $ v\in \mathcal{E}(M)$ and by \eqref{eq:EvaModPF}, $V(Y,q) = 2 \lambda^G$, where $\lambda^G$ relates to the eigenvalue $\lambda^M$ associated to $v$ as given by \eqref{eq:eigGM}. Additionally, since $0 < \Psi_M (Y) \leq 2 \lambda_{\max}^G$, it can be shown from \eqref{eq:WarpAngle_MPF} that 
    \begin{equation} \label{eq:WarpAngle_MPF_rge}
        0 < \theta (Y) < \pi/2 ,\ \forall X \in \mathrm{SO(3)}.
    \end{equation}
    In view of \eqref{eq:EvaModPFPi} and \eqref{eq:WarpAngle_MPF}, we can obtain 
    \begin{align}
        2(\lambda_{\max}^G)^2 ( 2 \lambda^G -\Psi_M(Y) ) & =  k^2 \Psi_M(Y)^2  \Delta(v,u_q) . \label{eq:Crit_V_MPF} 
    \end{align}
    The formula provides the solution $\Psi_M(Y)$. 
    
    We proceed to compute $\bar{\delta}_q$ given by \eqref{eq:bar_delta_q}. Let $h(v,u_q,u_p) \coloneqq 2 \sin^2( {\theta_{pq}}/{2}) \Delta(v,u_{pq}) $. By \eqref{eq:EvaModPFPi} and \eqref{eq:refined_Syn_gap}, $\bar{\delta}_q$ is written as 
    \begin{align}
        \bar{\delta}_q & = \min_{v \in \mathcal{E}(M)} 
        \Bigl(\max_{u_p \in \mathcal{U}_q } h(v,u_q,u_p) \Bigr),\label{eq:bar_delta_pf}
    \end{align}
    where $\theta_{pq}$ and $u_{pq}$ are given by \eqref{eq:u_pq_def} and \eqref{eq:theta_pq_def}.

    \setcounter{case}{0}
    \begin{case}[$\lambda_i^M = \lambda >0, \ i  = 1,2,3 $]
        The sets of warping directions $\mathcal{U}$ and $\mathcal{U}_q$ for each $q \in \mathcal{Q}$ are given in the item (1) in Theorem \ref{thm:SelectDirctions}. Note that $\mathcal{E}(M) = \mathbb{S}^2$ and $\lambda^G = 2 \lambda$. 
        
        Let us compute \eqref{eq:bar_delta_pf} in this case. Without loss of generality, we set $u_q = v_1$ and denote the eigenvector of $M$ by $v = \alpha_1 v_1 + \alpha_2 v_2 + \alpha_3 v_3$, $\sum_{i=1}^3\alpha_i^2 = 1 $. Hence, $\mathcal{U}_q = \{\pm v_2, \pm v_3 \}$ by Theorem \ref{thm:SelectDirctions}, and the associated $u_{pq}$ is given by \eqref{eq:u_pq_perp}. Additionally, we have $\Delta (v,u_q) =   2 \lambda \alpha_1^2$ from \eqref{eq:EvaModPFPi} and thus compute $\theta (Y)$ from \eqref{eq:Crit_V_MPF} and \eqref{eq:WarpAngle_MPF} as 
        \begin{align}
            \sin(\frac{\theta(Y)}{2}) = \frac{2k}{1 + \sqrt{1+4k^2 \alpha_1^2}}. \label{eq:sol_thetaY_1}
        \end{align}
        where $0 < k < \frac{\sqrt{2}}{2}$ since $\xi = 1$. It is obvious that $\theta(Y)$ are determined by $\alpha_1$. Using \eqref{eq:u_pq_perp}, \eqref{eq:theta_pq_perp}, and \eqref{eq:sol_thetaY_1}, we can compute \eqref{eq:bar_delta_pf} as
        \begin{align*}
            \bar{\delta}_q & = 4\lambda  \min_{\alpha_1^2 \leq 1} \biggl(  \sin (\frac{\theta(Y)}{2})   \Bigl( | \alpha_1  | \cos (\frac{\theta(Y)}{2}) + \sqrt{ \frac {1-\alpha_1^2}{2}}  \Bigr)   \biggr)^2 
        \end{align*}
        where we used \eqref{eq:WarpAngle_MPF_rge}, $\max \{(a+b)^2, (a-b)^2\} = (|a|+|b|)^2$ for $a,b \in \mathbb{R}$, and $\max \{|\cos(t)|,|\sin(t)|\} \geq \frac{\sqrt{2}}{2}$ for $t \in \mathbb{R}$. 
        
        Consider the function $\gamma : [0,1] \to \mathbb{R}_{\geq 0}$ defined as
        \begin{equation*}
            \gamma (\alpha_1) =\frac{1}{2} \alpha_1 \sin (\theta(Y))   +  \sin (\frac{\theta(Y)}{2}) \sqrt{ \frac {1-\alpha_1^2}{2}}  ,
        \end{equation*}
        where $\sin(\theta(Y)/2)$ is given by \eqref{eq:sol_thetaY_1}. One can verify that for $0 < k < \frac{\sqrt{2}}{2}$, $\frac{1}{2} \alpha_1 \sin (\theta(Y))  > \alpha_1 \gamma(1)$ and $\sin (\frac{\theta(Y)}{2}) \sqrt{ \frac {1-\alpha_1^2}{2}} > (1 - \alpha_1) \gamma(0)$ hold for all $\alpha_1 \in (0,1)$. Hence, $\gamma$ is strictly concave and $\min_{\alpha_1 \in [0,1]} \gamma (\alpha_1) = \min \{ \gamma (0), \gamma (1)\}$. Note that $\bar{\delta}_q =4 \lambda \min_{\alpha_1 \in [0,1]} \gamma^2 (\alpha_1)$. One can obtain \eqref{eq:refined_sy_gap_1eigx} by replacing with $\gamma^2 (0) = \frac{1}{2} k^2$ and $\gamma^2 (1) = \Xi_1^2  (1 - \Xi_1^2) $. This proves the item (1).
    \end{case}

    \begin{case}[$\lambda_1^M = \lambda_2^M > \lambda_3^M > 0 $]
      The sets $\mathcal{U}$ and $\{\mathcal{U}_q\}_{q \in \mathcal{Q}}$ are given by using item (2) in Theorem \ref{thm:SelectDirctions}. Note that $\mathcal{E}(M) =\{  v_3, - v_3  \} \cup \{v_{12} = v_1 \cos (t) + v_2 \sin(t): t \in \mathbb{R} \}  $, $\lambda^G_{\max} = \lambda_3^G$ and the gain $k$ satisfies $0 < k < 1/\sqrt{6 - 4 \xi^2} $ on account of $\xi > \frac{1}{2}$. 
      
      Let us compute \eqref{eq:bar_delta_pf} in two cases: $v = v_3$ or $v = v_{12}$. Without loss of generality, suppose $u_q = v_1$ and by Theorem \ref{thm:SelectDirctions}, $ \mathcal{U}_q = \{\pm v_2\}$.

      We first consider $v = v_3$. In view of \eqref{eq:EvaModPFPi}, one has $\Delta (v_3, v_1)  = \lambda_3^G - \lambda_2^G$ and thus computes $\theta(Y)$ from \eqref{eq:Crit_V_MPF} and \eqref{eq:WarpAngle_MPF} as 
      \begin{align}
        \sin (\frac{\theta(Y)}{2})  = \frac{2k }{1 + \sqrt{1 + 4 k^2 (1 - \xi  )}}. \label{eq:sol_theta_2_v3}
      \end{align}
      Making use of \eqref{eq:u_pq_perp}, \eqref{eq:theta_pq_perp}, \eqref{eq:u_pq_dot_v3_perp}, \eqref{eq:EvaModPFPi}, and $\Xi_{21} = \sin(\theta(Y)/2)$ in \eqref{eq:refined_sy_gap_2eigx}, one can obtain that
      \begin{align*}
        \max_{u_p \in \mathcal{U}_q } h(v_3,u_q,u_p)   =  2 \lambda_3^G \Xi_{21}^2 (1 + (1-2\xi)(1-\Xi_{21}^2) ).
      \end{align*}

      Second, let us turn to $v = v_{12}$, where $v_{12} = v_1 \cos (t) + v_2 \sin (t)$ for each $t \in \mathbb{R}$. By \eqref{eq:EvaModPFPi}, one has $\Delta (v_{12}, v_1)  = \lambda_2^G - \lambda_3^G \sin^2(t)$ and thus obtains from \eqref{eq:Crit_V_MPF} and \eqref{eq:WarpAngle_MPF} that
      \begin{align}
        \sin (\frac{\theta(Y)}{2})  =   \frac{2k \xi }{1 + \sqrt{1 + 4 k^2 \xi (\xi -  \sin^2 (t))}}. \label{eq:sol_theta_2_v12}
      \end{align}
      Note that the other solution to the quadratic equation \eqref{eq:Crit_V_MPF} is discarded for the restriction of $0< \sin (\frac{\theta(Y)}{2}) \leq k < 1$. Furthermore, combining \eqref{eq:u_pq_perp}, \eqref{eq:theta_pq_perp}, \eqref{eq:u_pq_dot_v3_perp}, \eqref{eq:u_pq_perp_perp}, and \eqref{eq:EvaModPFPi}, it follows that 
      \begin{align*}
        &\min_{v_{12}} \max_{u_p \in \mathcal{U}_q } h(v_{12},u_q,u_p) \\
        &\quad = \min_{t \in \mathbb{R}} \sin^2 (\theta(Y)) \left( \lambda_2^G - \frac{\lambda_3^G}{2}(1 - |\sin (2t)| \right) \\
        & \quad = 2 \lambda_3^G  \Xi_{22}^2 (1 - \Xi_{22}^2) (2\xi - 1)
      \end{align*}
      where the last equality is from the fact that the positive functions $\sin^2 (\theta(Y)) $ and $2 \lambda_2^G - \lambda_3^G (1 - |\sin (2t)|) > 0$ are both found the minimum value at $t = n \pi, \ n \in \mathbb{Z}$.

      Combining the above cases yields \eqref{eq:refined_sy_gap_2eigx}, which shows the item (2).
    \end{case}
    
    \begin{case}[$\lambda_1^M = \lambda^M_2 > \lambda^M_3 \geq 0$]
      The condition of this case is almost same as that of case 2 except that $\lambda_2^G$ can equal to $\frac{1}{2} \lambda_3^G$. Note that the sets of warping directions $\mathcal{U}$ and $\mathcal{U}_q$ for each $q \in \mathcal{Q}$ are given in item (3) in Theorem \ref{thm:SelectDirctions}. Without loss of generality, we can assume $u_q = v_1$ and thus obtain $\mathcal{U}_q = \{-v_1, v_{+s}, v_{-s}\}$ where $v_{\pm s} = \frac{1}{2} v_1 \pm \frac{\sqrt{3}}{2} v_2 $. The associated $u_{pq}$ and $\theta_{pq}$ are given by \eqref{eq:u_pq_pi3} and \eqref{eq:theta_pq_pi3}, respectively. Similar to case 2, we will compute \eqref{eq:bar_delta_pf} in two cases: $v = v_3$ or $v = v_{12}$.

      First, let us consider $v = v_3$. Making use of \eqref{eq:u_pq_pi3}-\eqref{eq:u_pq_dot_v3_3pi} and \eqref{eq:EvaModPFPi}, we have
      \begin{align*}
        h(v_3,v_1,-v_1) 
        &= 2 \lambda_3^G \sin^2 (\theta(Y)) (1-\xi) ,\\
        h(v_3,v_1,v_{\pm s}) 
        &= \frac{\lambda_3^G}{2}   \sin^2 (\frac{\theta(Y)}{2}) \Bigl( 3 + (1 - 4\xi) \cos^2 (\frac{\theta(Y)}{2}) \Bigr) 
      \end{align*}
      where $\sin (\theta(Y)/2)$ is given by \eqref{eq:sol_theta_2_v3}. Therefore, it follows 
      \begin{align*}
        \max_{u_p \in \mathcal{U}_q } h(v_3,u_q,u_p) = &  \lambda_3^G \max \Bigl\{ 8  \Xi_{21}^2 (1 - \Xi_{21}^2) (1 - \xi),\\
        & \frac{1}{2}  \Xi_{21}^2 (3 + (1-4\xi)(1-\Xi_{21}^2) ) \Bigr\}.
      \end{align*}

      Second, let us consider $v = v_{12}$ for $v_{12} = v_1 \cos (t) + v_2 \sin (t)$. In view of \eqref{eq:u_pq_pi3}-\eqref{eq:u_pq_perp_3pi} and \eqref{eq:EvaModPFPi}, $h(v_{12},u_q,u_p)$ is computed as
      \begin{align*}
        h(v_{12},v_1,-v_1) 
        &= 2 \lambda_3^G \sin^2 (\theta(Y)) (\xi -  \sin^2(t)) , \\
        h(v_{12},v_1,v_{\pm s}) &= \frac{\lambda_3^G}{2} \sin^2 (\theta(Y)) \left(\xi -  \sin^2 (t \pm \frac{\pi}{3})  \right),  
      \end{align*}
      where $\sin (\frac{\theta(Y)}{2})$ is given by \eqref{eq:sol_theta_2_v12}. As forth in case 2, the minima of $\sin^2 (\theta(Y))$ is found at $t = n \pi, n \in \mathbb{Z}$. Therefore, invoking Proposition \ref{prop:min_F} in Appendix, we have 
      \begin{align*}
        & \min_{v_{12}} \max_{u_p \in \mathcal{U}_q } h(v_{12},u_q,u_p)  \geq  2 \lambda_3^G  \Xi_{22}^2 (1 - \Xi_{22}^2) (\xi - \frac{1}{4}).
      \end{align*}
    
      Substituting these to \eqref{eq:bar_delta_pf} yields \eqref{eq:refined_sy_gap_2eigx1}. This shows the item (3). This finishes the proof. \qed

    \end{case} 
\end{pf}

\section{Proof of Proposition \ref{prop:stability}}
\begin{pf}
  With the function $V \in \mathcal{C}^1 (\mathrm{SO(3)} \times \mathcal{Q},\mathbb{R})$ centrally synergistic relative to $\mathcal{A}$, we consider the Lyapunov function candidate $U : \mathcal{X}_1 \to \mathbb{R}$ given by $U(x_1) = k_1 V(\tilde{R},q) +  \tilde{\omega}^\top J \tilde{\omega}$. 
  We can see that $U$ is positive definite relative to $\mathcal{A}_1$. The time derivative of $U$ along the flows is given by $\dot{U} =   - 2 k_2 | \tilde{\omega} |^2 $, where we used the facts that $\tilde{\omega}^\top \Sigma ( \tilde{\omega} , \omega_d)\tilde{\omega} = 0 $ for all $\tilde{\omega}, \omega_d \in \mathbb{R}^3$ and that $\operatorname{tr} (A^\top x^\wedge ) = 2 x^\top \psi (A)$ for all $A \in \mathbb{R}^{3 \times 3} $ and $u \in \mathbb{R}^3$. It shows that $\dot{U}(x_1) \leq 0$ for all $x_1 \in \mathcal{F}_1$. In addition, over the jumps, for each $x_1 \in \mathcal{J}_1$, we have that $U(x_1) - U(x_1^+) = k_1 \mu_V (\tilde{R},q) \geq k_1 \pi_V (\tilde{R},q) \geq k_1 \delta (q) > 0 $. Therefore, for the hybrid closed-loop system $\mathcal{H}_1$ given by \eqref{eq:hyb_clp1}, the set $\mathcal{A}_1$ is stable by \cite[Theorem 23]{Goebel2009}, the number of jumps is bounded by $U(x_{1}(0,0)) / ( k_{1} \min_{q \in \mathcal{Q}} \delta(q))$, and the solution $x_{1} (t,j)$ is bounded for all $(t,j) \geq (0,0)$. Therefore, the number of jumps is finite and the Zeno solution is avoided. Furthermore, invoking the hybrid invariance principle \cite[Theorem S13]{Goebel2009}, we must have that the solution to $\mathcal{H}_{1}$ approaches the largest weakly invariant set in $\dot{U}^{-1} (0) \coloneqq \{x_{1} \in \mathcal{F}_{1} : \dot{U} ({x_{1}}) = 0\}$. For each $x_1 \in \dot{U}^{-1} (0) $, we have $\tilde{\omega} \equiv 0$ and hence $\rho_V (\tilde{R},q ) \equiv 0$ from the dynamics of $\tilde{\omega}$ given by \eqref{eq:hyb_clp1} and \eqref{eq:ctr_law1}. It implies that $ \dot{U}^{-1} (0)  \subseteq \mathcal{I} \coloneqq \{ x_1 \in \mathcal{X}_1 : (\tilde{R},q) \in \operatorname{Crit} (V), \tilde{\omega} = 0 \}$. Since $V$ is centrally synergistic relative to $\mathcal{A}$, it follows by Lemma \ref{lem:cenSynergyLocal} that $\pi_V (\tilde{R},q) > \delta (q)$ for all $x_1 \in \mathcal{I} \setminus \mathcal{A}_1$. In consequence, $\mathcal{I} \cap \mathcal{F}_1 = \mathcal{A}_1 = \dot{U}^{-1} (0) $. Since $\mathcal{A}_1$ is invariant, we can conclude that $\mathcal{A}_1$ is globally attractive and thus globally asymptotically stable for the closed-loop system $\mathcal{H}_{1}$. Finally, we can assert the robustness of asymptotic stability to measurement disturbances by \cite[Theorem 3.26]{Sanfelice2021}, which completes the proof. \qed
\end{pf}

\section{Non-centrally synergistic hybrid controller}
Let $b_1, b_2 \in \mathbb{S}^2$ be the constant unit vectors expressed in the body-fixed frame such that $b_1^\top b_2 = 0$. For example, $b_1 = (1, 0, 0)^\top$, $b_2 = (0,1,0)^\top$. Let $\mathcal{Q} = \{1,2,3\}$. Define the error functions on $\mathrm{SO(3)} $ as $\Psi_{N_i} (X) = 1- \langle b_i , Xb_i \rangle  $, $\Psi_{E_i} (X)  = \alpha + \beta \langle b_i , X b_1^\wedge b_2 \rangle$, $i\in \{1,2\}$, where $\alpha,\beta$ are constant satisfying $1 < \alpha < 2$ and $|\beta| < \alpha - 1$. The corresponding error vectors are given by $e_{N_i} (X) = (Xb_i)^\wedge b_i$ and $e_{E_i} (X) = \beta b_i^\wedge  X (b_1^\wedge b_2)$ for $i\in \{1,2\}$. With these notations, the synergistic function $V: \mathrm{SO(3)} \times \mathcal{Q} \to \mathbb{R}$ is defined as $V (X,1) =  \Psi_{N_1} (X) + \Psi_{N_2} (X)$ and $V (X,q) =  \Psi_{N_{q-1}} (X) + \Psi_{E_{4-q}} (X)$ for $q = 2,3$. In addition, the hybrid error vector $e_H : \mathrm{SO(3)} \times \mathcal{Q} \to \mathbb{R}^3 $ is defined as $e_H (X,1) =  e_{N_1} (X) + e_{N_2} (X)$ and $e_H (X,q) =  e_{N_{q-1}} (X) + e_{E_{4-q}} (X)$ for $q = 2,3$. Let $0 < \delta < \min \{2- \alpha, \alpha - |\beta| - 1\}$. The flow and jump sets are given by $\mathcal{F}_2  = \left\{ (X,q) \in \mathrm{SO(3)} \times \mathcal{Q} : \mu_V(X,q) \leq \delta \right\}$ and $\mathcal{J}_2  = \left\{ (X,q) \in \mathrm{SO(3)} \times \mathcal{Q} : \mu_V(X,q) \geq \delta \right\}$. Finally, the synergistic hybrid controller in \cite{Lee2015} is given by
\begin{align}
  &
  \begin{cases}
    \begin{aligned}
      \dot{q} &= 0 & (R^\top R_d, q) \in \mathcal{F}_{2} ,\\
      q^+ &\in {\operatorname{argmin}}_{p \in \mathcal{Q}} V (R^\top R_d,p) &  (R^\top R_d, q) \in \mathcal{J}_{2},
    \end{aligned}
    \\
    \begin{aligned}
      \tau &= -k_1 e_H (R^\top R_d, q) - k_2 (\omega - R^\top R_d \omega_d ) \\
      & \quad + (R^\top R_d \omega_d)^\wedge J R^\top R_d \omega_d + J R^\top R_d \dot{\omega}_d .
    \end{aligned}
  \end{cases} \label{eq:ctr_nonCS}
\end{align}

\bibliographystyle{ieeetr}        
\bibliography{References}

\end{document}